# Redox-controlled conductance of polyoxometalate molecular junctions.

Cécile Huez,[1] David Guérin,[1] Stéphane Lenfant,[1] Florence Volatron,[2] Michel Calame,[3,4] Mickael L. Perrin,[3,5] Anna Proust[2] and Dominique Vuillaume.[1]

1) Institute for Electronics Microelectronics and Nanotechnology (IEMN), CNRS, University of Lille, Av. Poincaré, Villeneuve d'Ascq, France .
2) Institut Parisien de Chimie Moléculaire (IPCM), CNRS, Sorbonne Université, 4 Place Jussieu, F-75005 Paris, France.
3) EMPA, Transport at the Nanoscale Laboratory, 8600 Dübendorf, Switzerland.
4) Dept. of Physics and Swiss Nanoscience Institute, University of Basel, Klingelbergstrasse 82, 4056 Basel, Switzerland.
5) Department of Information Technology and Electrical Engineering, ETH Zurich, 8092 Zurich, Switzerland.

**Abstract.**

We demonstrate the reversible *in situ* photoreduction of molecular junctions of phosphomolybdate $[PMo_{12}O_{40}]^{3-}$ monolayer self-assembled on flat gold electrodes, connected by the tip of a conductive atomic force microscope. The conductance of the one electron reduced $[PMo_{12}O_{40}]^{4-}$ molecular junction is increased by ∼ 10, this open-shell state is stable in the junction in air at room temperature. The analysis of a large current-voltage dataset by unsupervised machine learning and clustering algorithms reveals that the electron transport in the pristine phosphomolybdate junctions leads to symmetric current-voltage curves, controlled by the lowest unoccupied molecular orbital (LUMO) at 0.6-0.7 eV above the Fermi energy with ∼25% of the junctions having a better electronic coupling to the electrodes than the main part of the dataset. This analysis also shows that a small fraction (∼ 18% of the dataset) of the molecules is already reduced. The UV light *in situ* photoreduced phosphomolybdate junctions are systematically featuring slightly asymmetric current-voltage behaviors, which is ascribed to electron transport mediated by the single occupied molecular orbital (SOMO) nearly at resonance with the Fermi energy of the electrode and by a closely located single unoccupied molecular orbital (SUMO) at ∼0.3 eV above the SOMO with a weak electronic coupling to the electrodes (∼ 50% of the dataset) or at ∼0.4 eV but with a better electrode coupling (∼ 50% of the dataset). These results shed lights to the electronic properties of reversible switchable redox polyoxometalates, a key point for potential applications in nanoelectronic devices.

**Introduction.**

A reliable knowledge of the relationship between the redox state of polyoxometalates (POMs) and their electron transport (ET) properties, e.g., electrical conductance, at the nanoscale is mandatory for potential applications in POM-based nanoelectronic devices. At the material level (macroscopic scale), an increase by a factor of about 500 in the conductivity of a single crystal of $HK_{3,5}Li_{1,5}Co_4[NaP_5W_{30}O_{110}]$ was reported upon ultraviolet photoreduction.[1] However, at the device level, the redox state of the POMs is determined and tuned with the POMs embedded in a dielectric (insulating) layer (gate dielectric of a transistor,[2] stacked layer of a capacitance[3-6]). In this case, the direct relationship between the redox state of the POMs and their electrical conductance cannot be studied. At the nanoscale, monolayers (or few monolayers) of POMs assembled on conducting surfaces by various approaches (covalent grafting, electrostatically immobilization, layer-by-layer deposition,…) and contacted by scanning probe microscopy have been used to measure the conductance of various POMs,[7-12] but the redox state was not systematically tuned and controlled. It is only recently that at the single molecule level, using scanning tunneling microscopy, Linnenberg et al. demonstrated a step-by-step increase of the POM (Lindqvist-type $V_6$) conductance up to the 4-electron reduced state.[13, 14]

Here, we study by conductive atomic force microscopy (C-AFM) the ET of a monolayer of an "archetype" Keggin-type POM ($[PMo_{12}O_{40}]^{3-}$) assembled on Au surfaces. We demonstrate that the redox switching of the POMs in the monolayer is triggered *in situ* by UV photoreduction and is reversible upon dark condition at room temperature (or moderate heating) and we report an increase of the POM conductance by a factor ≈ 10 upon a one-electron reduction. This conductance switching is ascribed to a transition from a LUMO-mediated ET for the pristine (fully oxidized) $PMo_{12}(0)$ to a SOMO- and SUMO-mediated ET for the one-electron reduced $PMo_{12}(I)$. The analysis of a large dataset (600 I-Vs measured by C-AFM) of the ET properties using unsupervised machine learning and clustering algorithms reveal that all the POM molecular junctions with the $PMo_{12}(0)$ display almost symmetric I-Vs, with nearly the same energy of the LUMO at around 0.6 - 0.7 eV above the Fermi energy of the electrodes and that about 25% of the molecular junctions have a better electronic coupling to the electrodes than the main part of the junctions in this dataset. The molecular junctions with the one-electron reduced $PMo_{12}(I)$ systematically display slightly asymmetric I-Vs (more current at a negative voltage applied on the Au substrate). The dataset is composed of two types of molecular junctions: one with the SUMO level at around 0.3 eV above the electrode Fermi level but with a weak electronic coupling to the electrodes (about half of the dataset), and the another one with the SUMO at about 0.4 eV but with a better electronic coupling to the electrodes. These features explain the large overall dispersion of the measured I-V dataset.

## Synthesis of molecules and self-assembled monolayers, physico-chemical characterizations.

The fully oxidized (pristine) phosphomolybdate $[PMo_{12}O_{40}]^{3-}$ has been prepared as a tetrabutylammonium ($N(C_4H_9)_4^+$ or $TBA^+$) salt as reported in the literature.[15] Some of us have previously described its mono-electronic reduction by reaction with phenyllithium, according to the following equation.[16]

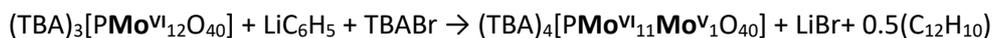

$(TBA)_3[PMo^{VI}{}_{12}O_{40}]$ + $LiC_6H_5$ + TBABr → $(TBA)_4[PMo^{VI}{}_{11}Mo^{V}{}_1O_{40}]$ + LiBr + $0.5(C_{12}H_{10})$

$(TBA)_3[PMo^{VI}{}_{12}O_{40}]$ ($PMo_{12}(0)$ for short) and $(TBA)_4[PMo^{VI}{}_{11}Mo^{V}{}_1O_{40}]$ ($PMo_{12}(I)$ for short) are thus available as yellow and blue powders, respectively. Relevant details about the characterization of the products ($^{31}P$ NMR, cyclic voltammetry, solution UV-vis spectra, XPS spectra on powder) are given in the Supporting Information (section 1). In particular, the cyclic voltammogram (Fig. S3), the UV-vis spectra of the POMs in solution (Fig. S4) and the XPS on powder (Mo $3d_{3/2}$ and $3d_{5/2}$ orbit doublet, Fig. S5) clearly show the reduction of the POMs.

For the monolayer preparation, we used ultra-flat template-stripped gold surface $^{TS}$Au.[17] The freshly prepared $^{TS}$Au surfaces were first functionalized with a 6-aminohexane-1-thiol hydrochloride (HS-$(CH_2)_6$-$NH_3^+$ / $Cl^-$) SAM (self-assembled monolayer), C6 SAM for short. Then these SAMs were treated by a PBS (phosphate-buffered saline, pH=7.4) solution for 2 hours, followed by ultra-sonication in DI water for 5 minutes to adjust the ratio of $NH_3^+$/$NH_2$ on the surface and optimize the next step (electrostatic deposition of the POMs) as reported in our previous work.[11] After this process, we estimate that ~ 30% - 50% of the amine terminal groups are protonated ($NH_3^+$)-(details in the Supporting Information).[18-21] Then, the $PMo_{12}$(0) and $PMo_{12}$(I) molecules were electrostatically attached on these positively charged C6-SAMs by dipping the modified substrate in the POM solution (1h at $10^{-3}$M in acetonitrile) and rinsed in acetonitrile, more details in the Supporting Information (section 2). The electrical neutrality is thus ensured by a mixture of the $TBA^+$ counterions and the $NH_3^+$ end-goups of the alkyl chains (see details in the Supporting Information).

The thickness of the SAMs was characterized by spectroscopic ellipsometry (see section 3 in the Supporting Information) at each step: 1.0 ± 0.2 nm for C6-SAM, 1.8 ± 0.2 nm for C6/$PMo_{12}$(0) and C6/$PMo_{12}$(I) SAMs. Tapping mode AFM images (Fig. 1) show that the C6-SAMs are flat with a rms roughness of ≈ 0.65 nm (reference for $^{TS}$Au : 0.3-0.4 nm) and free of defects (no pinhole, nor aggregate, the dark spots are defects (pinhole) in the underlying Au substrate and they are masked for the roughness analysis). After the deposition of the POMs, the surface is still featureless with a rms roughness of ≈ 0.56 nm for $PMo_{12}$(0) and ≈ 0.73 nm for PMo$_{12}$(I). These results indicate that the C6-SAMs are densely packed (the theoretical thickness for a fully packed SAM with the C6 alkyl chains almost perpendicular to the substrate is *ca.* 1.2 nm). The average thickness for the POM part is 0.8 nm, slightly smaller than the nominal size of the POM (1 nm), which can be explained considering the voids in a close-packed monolayer of spheres.[22] We conclude that the POM coverage is almost complete.

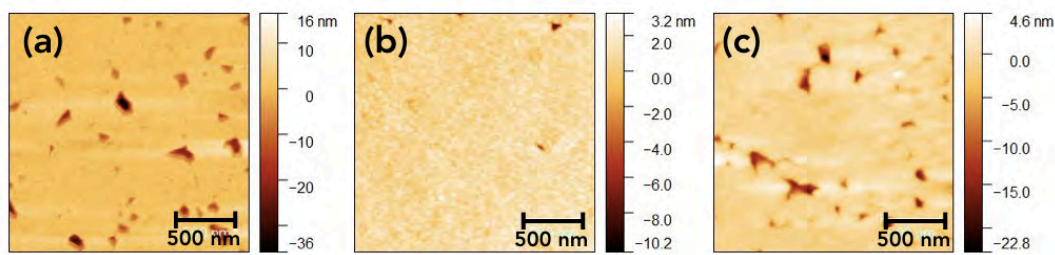

*Figure 1*. Tapping-Mode AFM images of (a) $^{TS}$Au-C6 SAM, (b) $^{TS}$Au-C6/PMo$_{12}$(0) and (c) $^{TS}$Au-C6/PMo$_{12}$(I). The dark spots are defects (pinhole) in the underlying Au substrate and they are masked for the roughness analysis. The rms roughness is 0.65 nm, 0.56 and 0.73 nm for the $^{TS}$Au-C6 SAM, $^{TS}$Au-C6/PMo$_{12}$(0) and $^{TS}$Au-C6/PMo$_{12}$(I) samples, respectively.

**Electron transport properties.**

The electron transport properties were measured by C-AFM for the two systems $^{TS}$Au-C6/PMo$_{12}$(0)//Pt and $^{TS}$Au-C6/PMo$_{12}$(I)//Pt ("-" denotes a chemical bond, "/" an electrostatic contact and "//" a mechanical contact). Typically up to 600 current-voltage (I-V) curves were acquired at several locations on the SAMs using a C-AFM tip (Pt tip, grounded), at a low loading force F ≈ 6-8 nN (see details section 4 in the Supporting Information). Figure 2 shows the 2D histograms (heat map) of hundreds of I-V traces for both the $^{TS}$Au-C6/PMo$_{12}$(0)//Pt and $^{TS}$Au-C6/PMo$_{12}$(I)//Pt junctions prepared directly with the pristine and reduced POMs. In these datasets, we removed the I-V traces reaching the saturating current of the preamplifier during the voltage scan (typically ~3-7% IVs of the dataset) and those close to the sensitivity limit of the C-AFM (very noisy) or with large and abrupt changes in the measured current (C-AFM tip contact issue), typically ~15-20% of the IVs (see section 4 in the Supporting Information (Fig. S4) and the machine learning/clustering analysis below and in section 7 in the Supporting Information). For the $^{TS}$Au-C6/PMo$_{12}$(0)//Pt sample a large number of I-V traces are low (<10$^{-11}$ A) and noisy compared to the $^{TS}$Au-C6/PMo$_{12}$(I)//Pt junction, which explains the smaller number of traces retained in the 2D histogram of the $^{TS}$Au-C6/PMo$_{12}$(0)//Pt sample. The black lines in Figs. 2a and 2b are the calculated mean current I-V curve (denoted as Ī-V). The current histograms at +1.5 V and -1.5 V are also shown in Figs. 2c and 2d, they are fitted by a log-normal distribution. The fitted parameters, the log-mean current (log-μ), the corresponding mean current Ī, and the log-standard deviation (log-σ) are summarized in Table 1. From these data, we clearly observe a larger increase of the current by a factor ≈ 10 for the PMo$_{12}$(I) molecules. We also observe a slight asymmetry of the I-Vs for the $^{TS}$Au-C6/PMo$_{12}$(I)//Pt, with a negative asymmetry ratio R$^-$ = Ī(-1.5 V)/Ī(+1.5 V) ≈ 5.5 (evaluated from the mean of current histograms, Figs. 2c and 2d).

|  |  | TSAu-C6/PMo$_{12}$(0)//Pt | TSAu-C6/PMo$_{12}$(I)//Pt |
|---|---|---|---|
| +1.5V | log-μ | -9.92 | -9.50 |
|  | Ī (A) | 1.2x10$^{-10}$ | 3.1x10$^{-10}$ |
|  | log-σ | 0.68 | 0.72 |
| -1.5V | log-μ | -9.70 | -8.80 |
|  | \|Ī\| (A) | 2.0x10$^{-10}$ | 1.6x10$^{-9}$ |
|  | log-σ | 0.43 | 0.36 |

*Table 1. Parameters of the log-normal fits of the current distributions at 1.5V and -1.5V (Figs. 2c and 2d): log-mean current (log-μ), the corresponding mean current Ī, and the log-standard deviation (log-σ).*

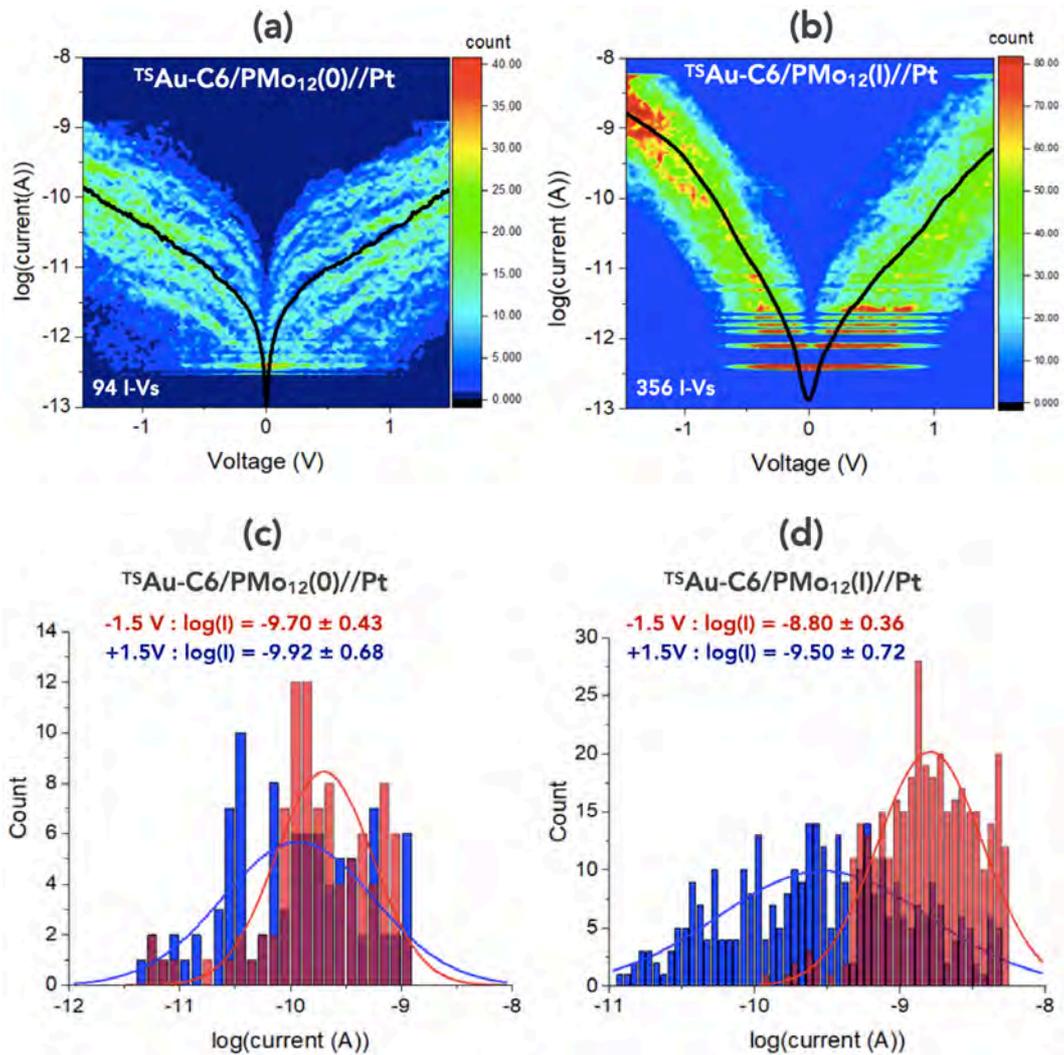

*Figure 2*. 2D histograms (heat maps) of I-Vs in a semi-log$_{10}$ plot for (a) $^{TS}$Au-C6/PMo$_{12}$(0)//Pt and (b) $^{TS}$Au-C6/PMo$_{12}$(I)//Pt. The solid black lines are the mean $\bar{I}$-V curves. The histograms of the currents at -1.5 V (red bars) and +1.5 V (blue bars) and fits by a log-normal distribution (the fitted log-mean value ± log-standard deviation is indicated on the panels).

We then follow *in situ* by C-AFM the reversible reduction/re-oxidation of the POMs in the SAMs. Starting from a $^{TS}$Au-C6/PMo$_{12}$(0)//Pt junction, we irradiate the POM layer with UV light (at 308 nm during few hours, see details sections 5 and 6 in the Supporting Information) and then we turn off the light and let the POM layer in ambient air either at room temperature (few tens of hours) or under a moderate heating (2h at 80°C on a hotplate in air). Both the POMs and the C6 SAMs are stable at this temperature (see in the Supporting Information). The I-Vs are measured after each step and the figure 3 shows the evolution the mean current (at -1 V) for 3 irradiation/relaxation cycles (Fig. S10 with more data). The complete dataset measured after each step is given in Fig. S10. A clear increase in the conductance is observed upon UV irradiation, followed by a return to a state of lower conductance after turning the light off. We note that the reduced state of the POMs in the junctions is stable long enough to do the C-AFM measurements (couple of hours) and that the return to the oxidized state laps for tens hours to a day (in air and at room temperature). From the experimental behavior of the monolayers directly prepared with the POMs in their PMo$_{12}$(0) and PMo$_{12}$(I) states described just above (Fig. 2), we can infer that the PMo12(0) monolayer is *in situ* photo-reduced and subsequently air re-oxidized. This is consistent with the reported photochemical reduction of polyoxometalates underlining their application in photochromic materials.[23] Irradiation in the UV range indeed results in oxygen-to-metal charge transfer transitions. This behavior was also checked on a drop cast film of PMo$_{12}$(0) on a glass substrate and irradiated with the same UV source (color change - see Fig. S11). We thus demonstrate that we are able to switch the redox state of the PMo$_{12}$ at the monolayer scale.

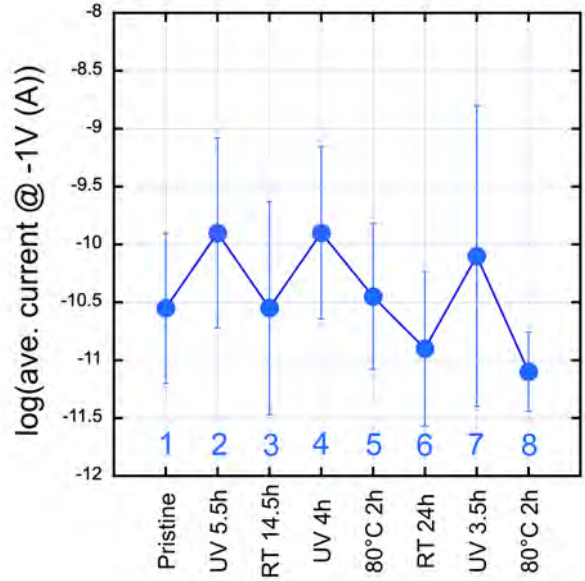

*Figure 3.* Evolution of the mean current $\log(|\bar{I}|)$ at -1V for 3 cycles of reduction/ oxidation. The currents measured during the last cycle are lower, possibly due to some degradation of the sample or the tip or a drift of the C-AFM loading force (see the Supporting Information).

To analyze the electronic structure in more detail, we fit all the individuals I-V curve in the dataset shown in Fig. 2 with the single-energy level (SEL) model (Eq. 1), considering that: i) a single molecular orbital (MO) dominates the charge transport, ii) the voltage mainly drops at the molecule/electrode interface and iii) that the MO broadening is described by a Lorentzian or Breit-Wigner distribution:[24, 25]

$$I(V) = N \frac{8e}{h} \frac{\Gamma_1 \Gamma_2}{\Gamma_1 + \Gamma_2} \left[ \arctan\left( \frac{\varepsilon_{0-SEL} + \frac{\Gamma_1}{\Gamma_1 + \Gamma_2} eV}{\Gamma_1 + \Gamma_2} \right) - \arctan\left( \frac{\varepsilon_{0-SEL} - \frac{\Gamma_2}{\Gamma_1 + \Gamma_2} eV}{\Gamma_1 + \Gamma_2} \right) \right]$$

(1)

with $\varepsilon_{0-SEL}$ the energy of the MO involved in the transport (with respect to the Fermi energy of the electrodes), $\Gamma_1$ and $\Gamma_2$ the electronic coupling energy between the MO and the electron clouds in the two electrodes, e the elementary electron charge, h the Planck constant and N the number of molecules contributing to the ET in the molecular junction (assuming independent molecules conducting in parallel, *i.e.* no intermolecular interaction[26-28]). Albeit this number can be estimated using mechanical models of the tip/SAM interface in some cases when the Young modulus of the SAM is reasonably known,[29-33] this is not the case here for the POM/alkyl SAM system for which the Young modulus has not been determined. Consequently, we use N=1 throughout this work. This means that the $\Gamma_1$ and $\Gamma_2$ values are "effective" coupling energies of the SAM with the electrodes and they are used only for a relative comparison of the POM SAMs measured with the same C-AFM conditions in the present work and they cannot be used for a direct comparison with other reported data (as for example from single molecule experiment). We also note that the exact value of N has no significant influence on the fitted parameter $\varepsilon_0$. We also used the transition voltage spectroscopy (TVS)[34-38] to analyze the I-V curves. Plotting $|V^2/I|$ vs. V (Fig. S7),[39] we determine the transition voltages $V_{T+}$ and $V_{T-}$ for both voltage polarities at which the bell-shaped curve is maximum. This threshold voltage indicates the transition between off-resonant (below $V_T$) and resonant (above $V_T$) transport regime in the molecular junctions and can therefore be used to estimate the location of the energy level. The value of $\varepsilon_{0-TVS}$ is estimated by:[36]

$$|\varepsilon_{0-TVS}| = 2 \frac{e|V_{T+}V_{T-}|}{\sqrt{V_{T+}^2 + 10|V_{T+}V_{T-}|/3 + V_{T-}^2}} \tag{2}$$

We also determined an average value of the electrode coupling energy $\Gamma_{TVS}$ using this relationship:[40, 41]

$$G(0) = NG_0 \frac{\Gamma_{TVS}^2}{\varepsilon_{0-TVS}^2} \tag{3}$$

with G(0) the zero-bias conductance, $G_0$ the conductance quantum ($2e^2$/h=7.75x10$^{-5}$ S, e the electron charge, h the Planck constant) and N the number of molecules in the junction. G is calculated from the slope of the I-V curve in its ohmic region (-50 mV/50 mV) and N=1 (see above). Note that $\Gamma_{TVS}$ is equivalent to the geometrical average of the SEL values $(\Gamma_1\Gamma_2)^{1/2}$.[40, 41]

We combined the fit of the SEL model with the TVS method and we limited the fits of the SEL model to a voltage window -1 V to 1 V to obtain the best determination of $\varepsilon_0$ (see in the Supporting Information, section 4, for details on the fit protocol, Fig. S8). Also note that the SEL model is valid as long as the applied voltage does not drive the MO near the Fermi energy of the electrodes (*i.e.*, ET in the off-resonance situation), which is not the case for the PMo$_{12}$(I). Consequently, this model is not used in that case (see section 4 in the Supporting Information), and the I-V measurements are only analyzed with the TVS method in this latter case. Figure 4 shows the statistical distribution of the $\varepsilon_{0-SEL}$ and $\varepsilon_{0-TVS}$ values obtained by fitting the SEL model and applying the TVS analysis on every I-V trace of the datasets (Fig. 2) for the C6-PMo$_{12}$(0) and C6-PMo$_{12}$(I) molecular junctions.

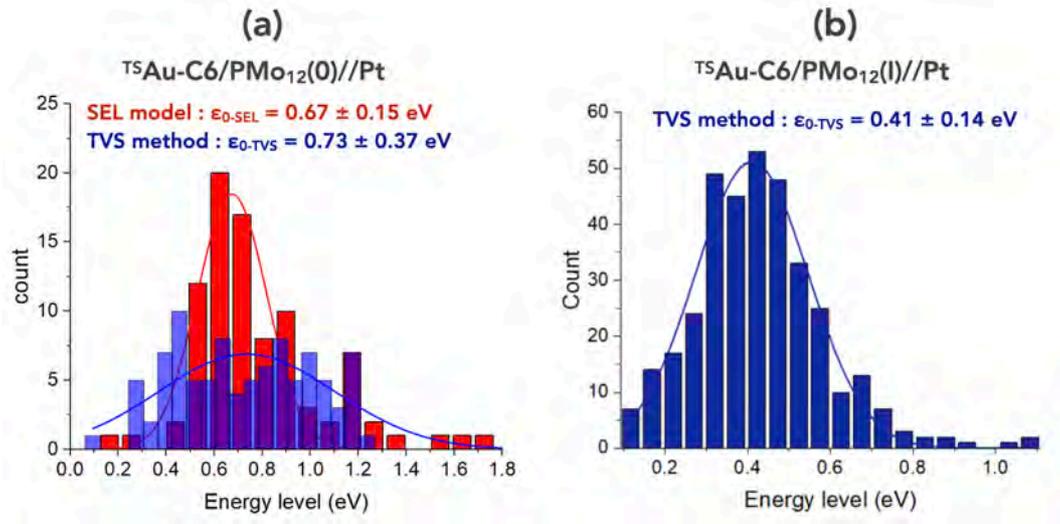

*Figure 4. Statistical distribution of the energy level (SEL model and TVS method) involved in the ET properties for (a) $^{TS}$Au-C6/PMo$_{12}$(0)//Pt and (b) $^{TS}$Au-C6/PMo$_{12}$(I)//Pt. The solid lines are the fits by a Gaussian distribution with the mean value ± standard deviation indicated in the panels.*

We conclude that the increase of the current for the $^{TS}$Au-C6/PMo$_{12}$(I)//Pt molecular junction is mainly due to the fact that MO involved in the electron transport comes closer to the Fermi level of the electrodes. In the case of the SEL model applied on the dataset of the $^{TS}$Au-C6/PMo$_{12}$(0)//Pt samples, the electrode coupling energies $\Gamma_1$ and $\Gamma_2$ are also broadly distributed (0.01 - 1 meV, Fig. S9). For the TVS method, the average values of $\Gamma_{TVS}$ determined using Eq. (3) are given in Table 2 and are on reasonable agreement with the SEL values.

|  |  | $^{TS}$Au-C6/PMo$_{12}$(0)//Pt | $^{TS}$Au-C6/PMo$_{12}$(I)//Pt |
|---|---|---|---|
| TVS | $\varepsilon_{0-TVS}$ (eV) | 0.73 ± 0.37 | 0.41 ± 0.14 |
| TVS | $\Gamma_{TVS}$ (meV) | 0.32 | 0.1 |
| SEL | $\varepsilon_{0-SEL}$ (eV) | 0.67 ± 0.15 | n.a. |
| SEL | $\Gamma_1$ (meV) | 0.1-0.2 | n.a. |
| SEL | $\Gamma_2$ (meV) | 0.1-0.2 | n.a. |

***Table 2.*** *Parameters of the Gaussian fits of the molecular energy level $\varepsilon_0$ for the TVS method and the SEL model (Fig. 4) and of the electrode coupling energies. For the electronic coupling to the electrodes ($\Gamma_1$ and $\Gamma_2$) a range is indicated due to a large dispersion (see Fig. S9 in the Supporting Information).*

To gain a more detailed understanding and because the I-V traces, and consequently the values of the energy levels and electrode coupling energies, are largely dispersed (Figs. 2 and 4), we apply machine learning (ML) and clustering tool to classify the individual I-V trace according to common characteristic features appearing in the dataset (pattern recognition).[42-45] More specifically, we use an unsupervised, reference-free tool developed by some of us.[44, 45] Following the benchmark reported in Ref. 45, we use: (i) the UMAP(cos.) (uniform manifold approximation and projection with a cosine distance metric) for the construction of the feature space, (ii) the GAL (graph average linkage) for the clustering algorithm with an optimal number of 5 clusters[45] (more detail section 7 in the Supporting Information). Figure 5a shows the feature space obtained for the dataset of the C6-PMo$_{12}$(0) molecular junctions (note that for this ML-based analysis, the complete dataset of 600 I-Vs is used). Figure 5b shows the mean Ī-V of the 5 clusters. The same data are shown for C6-PMo$_{12}$(I) in Figs. 5c and 5d.

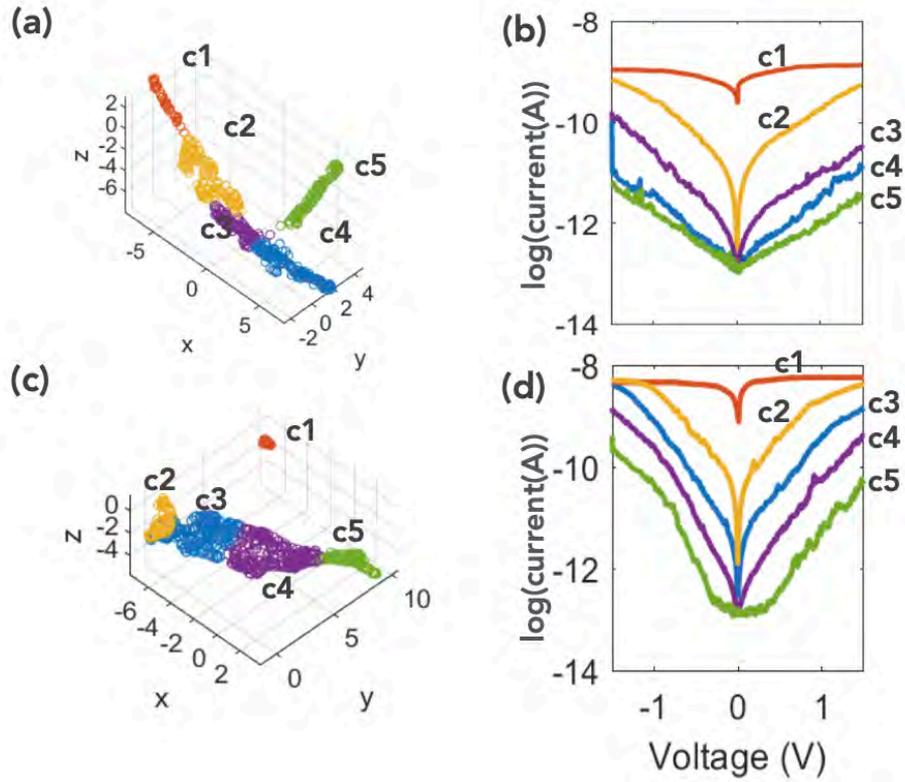

*Figure 5*. *(a) Feature space and (b) mean Ī-V for each clusters c1 to c5 for the $^{TS}$Au-C6-PMo$_{12}$(0)//Pt junctions. (c) Feature space and (d) mean Ī-V for each clusters c1 to c5 for the $^{TS}$Au-C6-PMo$_{12}$(I)//Pt junctions.*

For the $^{TS}$Au-C6-PMo$_{12}$(0)//Pt junctions, cluster 1 corresponds to I-Vs (7.7% of the dataset) saturating during the measurements, the mean Ī-V curve of clusters 2 to 5 shown in Fig. 5b were analyzed with the SEL models and the TVS approach (the complete datasets of I-Vs belonging to each cluster, the SEL fits and TVS curves are given in the Supporting Information (Figs. S14-S16). Cluster 2 (24.6%) and cluster 3 (18.8%) are characterized by the same $\varepsilon_0$ value (0.60-0.61 eV by TVS, 0.67-0.69 eV by SEL) with a higher electronic coupling to the electrodes for cluster 2 ($\Gamma_{TVS}\approx0.63$ meV; $\Gamma_1\approx0.44$ meV and $\Gamma_2\approx0.37$ meV by SEL) compared to cluster 3 ($\Gamma_{TVS}\approx0.14$ meV; $\Gamma_1\approx0.11$ meV and $\Gamma_2\approx0.083$ meV by SEL), data summarized in Table 3. Cluster 4 (29.2%) and cluster 5 (19.7%), with the lowest current and almost similar mean Ī-V traces (Fig. 5b) are, however, characterized by slightly different couple of the $\varepsilon_0$ and $\Gamma_{TVS}$ parameters that counterbalance each other (larger $\varepsilon_0$ for cluster 5 with a better electrode coupling : $\varepsilon_{0-TVS} \approx 0.72$ eV, $\Gamma_{TVS} \approx 0.28$ meV for cluster 5 vs. $\varepsilon_{0-TVS} \approx 0.55$ eV, $\Gamma_{TVS} \approx 0.054$ meV for cluster 4 (Table 3) (same behavior for the data obtained by SEL, Table 3). A larger $\varepsilon_0$, i.e. a MO far away from the Fermi energy tends to decrease the current, while a better electrode coupling energy tends to increase the current. We also note that clusters 2, 4 and 5 display almost symmetric mean Ī-Vs (with $R^- = \bar{I}(-1.5\ V)/(1.5\ V) < 2$), while a slight negative asymmetry is observed for cluster 3 with $R^- \approx 2.8$ (Table 3). This trend is confirmed by a statistical analysis on all the I-Vs belonging to each clusters (Fig. S7). For the $^{TS}$Au-C6-PMo$_{12}$(I) devices, cluster 1 (2.8% of the I-V traces) concerns again saturating I-Vs, which are not analyzed. The mean Ī-V curves for the 4 other clusters are analyzed by the TVS method (Figs. S20). The clusters 2 and 3 are characterized by a $\varepsilon_{0-TVS}$ value around 0.4 eV and $\Gamma_{TVS}$ 0.35 - 1.1 meV (Table 3), while clusters 4 and 5 have a lower $\varepsilon_{0-TVS}$ around 0.3 eV but also a lower $\Gamma_{TVS}$ (0.068-0.094 meV). In the case of the reduced POM, we note that all the mean Ī-Vs traces are asymmetric with the clusters 4 and 5 displaying the highest $R^-$ values (5-8, Table 3).

|  | ^TSAu-C6/PMo$_{12}$(0)//Pt | | | | ^TSAu-C6/PMo$_{12}$(I)//Pt | | | |
|---|---|---|---|---|---|---|---|---|
| cluster | C2 (24.6%) | C3 (18.8%) | C4 (29.2%) | C5 (19.7%) | C2 (12.5%) | C3 (35.7%) | C4 (35.5%) | C5 (13.5%) |
| $\varepsilon_{0\text{-TVS}}$ (eV) | 0.61 | 0.60 | 0.55 | 0.72 | 0.43 | 0.38 | 0.33 | 0.28 |
| $\Gamma_{TVS}$ (meV) | 0.63 | 0.14 | 0.054 | 0.28 | 1.1 | 0.35 | 0.094 | 0.068 |
| $\varepsilon_{0\text{-SEL}}$ (eV) | 0.69 | 0.67 | 0.60 | 0.81 | n.a. | | | |
| $\Gamma_1$ (meV) | 0.44 | 0.11 | 0.038 | 0.046 | | | | |
| $\Gamma_2$ (meV) | 0.37 | 0.083 | 0.036 | 0.032 | | | | |
| $(\Gamma_1.\Gamma_2)^{1/2}$ | 0.40 | 0.096 | 0.037 | 0.038 | | | | |
| R$^-$ | ≈1.7 | ≈2.8 | ≈1.1 | ≈1.5 | ≈2* | ≈2.6 | ≈5 | ≈8 |

*Table 3. Parameters of the molecular energy level $\varepsilon_0$ for the TVS method and the SEL model of the mean Ī-V curves belonging to the different clusters (Fig. 5) and of the electrode coupling energies (detailed data shown in Figs. S10-S15 in the Supporting Information). R$^-$ is the asymmetry ratio, R$^-$ = Ī(-1.5 V)/Ī(1.5 V), calculated from the mean Ī-Vs of each cluster. * stands for underestimated value since the current saturates below around -1V in that case.*

**Discussion.**

From the I-V analysis (both on the mean Ī-V, Figs. 2 and 4, and the statistical measurements, Figs. S18 - S23) we propose the energy scheme shown in Fig. 6 for the ^TSAu-C6/PMo$_{12}$(0)//Pt and ^TSAu-C6/PMo$_{12}$(I)//Pt molecular junctions. For the PMo$_{12}$(0), the ET is mediated by the LUMO that is at about 0.7 eV above the Fermi energy ($\varepsilon_{0\text{-SEL}}$= 0.67 ± 0.15 eV, $\varepsilon_{0\text{-TVS}}$ = 0.73 ± 0.37 eV, Table 2) as determined by the SEL and TVS analysis of the I-V measurements (Fig. 6a). This value is consistent with a LUMO at -4.5 eV (*vs.* vacuum energy, theory)[46] and work functions (WF) of Au (≈4.8-5.2 eV) and Pt (≈5.6 eV), the exact alignment of the MOs with the Fermi energy of the electrodes being dictated by the interface dipole and the details of the molecule/metal contact[47-49]. We note that this LUMO energy level is also consistent with the cyclic voltammetry measurements (LUMO at ~ -4.9 eV, see Fig. S3). According to previously reported calculations, the HOMO is located deeper (HOMO-LUMO gap of ca. 1.9-2.2 eV)[46, 50] and it is not involved in the ET property of the molecular junction. For the one-electron reduced $PMo_{12}(I)$, the added electron is localized on a SOMO located near to the electrode Fermi level as also reported when open-shell radicals are incorporated in molecular junctions.[51-56] Generally speaking, it is known from ab-initio calculations that the SOMO-SUMO gap of a series one-electron reduced POMs is low at around ~0.2 eV,[57] and thus both levels can now easily contribute to ET in the molecular junctions, leading to the experimentally observed enhanced conductance (Fig. 6b). Consequently, we ascribe the experimental value of $\varepsilon_{0-TVS}$ = 0.41 ± 0.14 eV (Table 2) to the SUMO level, while the SOMO remains close to the Fermi energy. This result pinpoints the good stability of the open-shell structure of the PMo(I) molecules when inserted in the junctions at room temperature, while previous attempts to incorporate organic radicals in molecular junctions showed a stable open-shell junction only at low temperature and/or under UHV.[55, 58, 59] This is likely because the SOMO is localized on the Mo and therefore it is embedded inside the molecule and partly protected from a too strong interaction with the metal electrodes in a similar way as recently demonstrated for a verdazyl radical, stable in its open-shell configuration at room temperature in molecular junctions.[56]

We note that I-V measurements on a reference sample without the POMs, i.e. $^{TS}$AU-C6//Pt junctions, give a higher value $\varepsilon_0$ ≈ 0.9 eV (section 8 in the Supporting Information), in good agreement with previous results for the LUMO of alkyl chains on Au.[60, 61] Thus the above determined values of $\varepsilon_0$ can be solely attributed to MOs of the POMs, the alkyl chain SAM playing the role of a thin tunnel barrier between the Au electrode and the POMs (in addition to a template structure to the electrostatic deposition of the POMs).

The difference between the ET through $PMo_{12}(0)$ and $PMo_{12}(I)$ is the shape of the I-Vs: almost symmetric for $PMo_{12}(0)$ and asymmetric for $PMo_{12}(I)$. Considering the energy diagram for the $^{TS}Au-C6/PMo_{12}(0)//Pt$ (Fig. 6a), the position of the LUMO near the grounded Pt C-AFM tip is likely to induce a slight negative asymmetry (*i.e.*, more current at negative voltages applied on the Au substrate when the Fermi energy of the Au electrode moves upper towards the LUMO level) and allows LUMO-assisted ET (blue arrow in Fig. 6a), while for V>0 the LUMO does not enter in the energy window defined by the Fermi energy of the two electrodes, Fig. 6a.[62-65] However, the difference in the work function (WF) of the electrodes (lower WF for Au than for Pt) can induce a reverse behavior, a positive asymmetry, *i.e.*, more direct tunneling current (green arrow in Fig. 6a) for a positive voltage applied on the electrode with the lowest WF (i.e. Au),[66] as observed on the I-V curves of the C6 SAMs (Fig. S24 in the supporting information) showing a slight positive asymmetry. Since the $^{TS}Au-C6/PMo_{12}(0)//Pt$ and the $^{TS}Au-C6//Pt$ junctions display current levels of the same order of magnitude (Fig. 2 and Fig. S24 in the Supporting Information), these two effects of opposite behavior can counterbalance each other, leading to the observed almost symmetric I-V behavior.

On the contrary, all the I-Vs of the dataset for the $^{TS}Au-C6/PMo_{12}(I)//Pt$ sample, display a negative asymmetry (Fig. S21). Since the SOMO of the open-shell POM is lying close to the Fermi energy of the electrodes and the SOMO-SUMO gap is small ($\simeq$ 0.4 eV), we hypothesize that at V<0 the two levels can be involved in the ET of the POM junction (Fig. 6b), while only the SOMO is considered at V>0 since it remains close to the Fermi energy and the SUMO is no longer contributing to the ET. We assume that the direct tunneling is negligible since the PMo$_{12}$(I) junctions have a current one decade higher than the C6 alone SAMs. Thus, the fact that two channels contribute to the ET at V<0 might explain this negative asymmetry. These simple hypothesis must be confirmed by more detailed *ab-inito* calculations, the exact I-V behavior of the molecular junctions being dependent on the shift of MO under the applied bias (Stark effect), and, moreover, the presence of many dipoles in the molecular junctions (*e.g.,* between (PMo$_{12}$)$^{4-}$ and NH$_3$$^+$, (PMo$_{12}$)$^{4-}$ and TBA$^+$, charge transfer at the electrodes) is likely to change the energy landscape in the junction[67] and consequently the ET properties, especially the direction of the asymmetry of the I-Vs.[68-70] Finally, we note that change of the shape of the I-V curves (asymmetric vs. symmetric) has also been recently observed when switching a molecule between open-shell (radical) and closed-shell configurations.[54, 56]

It is also likely that the details on how these dipoles are organized (ordered vs. disordered), where they are located in the SAM, as well as the molecular ordering of the POMs influence the large dispersion of the I-V measurements observed here. This feature calls for more experiments, *e.g.* UHV-STM measurements with a molecular resolution, and AC-STM[71] allowing dielectric spectroscopy at the molecular level,[72] which are out of the scope of this work.

As noted the electrical neutrality in the SAM is ensured by a mixture of TBA$^+$ and NH$_3$$^+$ ions, which can also influence the ET in the molecular junction. It has been calculated that the presence of the positive counterions lowers the energy of the LUMO of the POMs. However the exact chemical nature of these ions has a negligible effect on this energy position.[73] Thus the TBA$^+$/NH$_3$$^+$ ratio (see Supporting Information) is not a crucial parameter. However, it was shown by the same authors that the presence of the counterions does not create additional conduction channels in the molecular junctions, but rather that they modify the potential landscape viewed by the POMs that transmit the electrons through the molecular junctions, and consequently can influence the I-V behavior. Thus, we surmise that a part of the large dispersion of the I-V curves might be due to local variations of this $TBA^+/NH_3^+$ ratio and how these counterions are organized (*vide supra*).

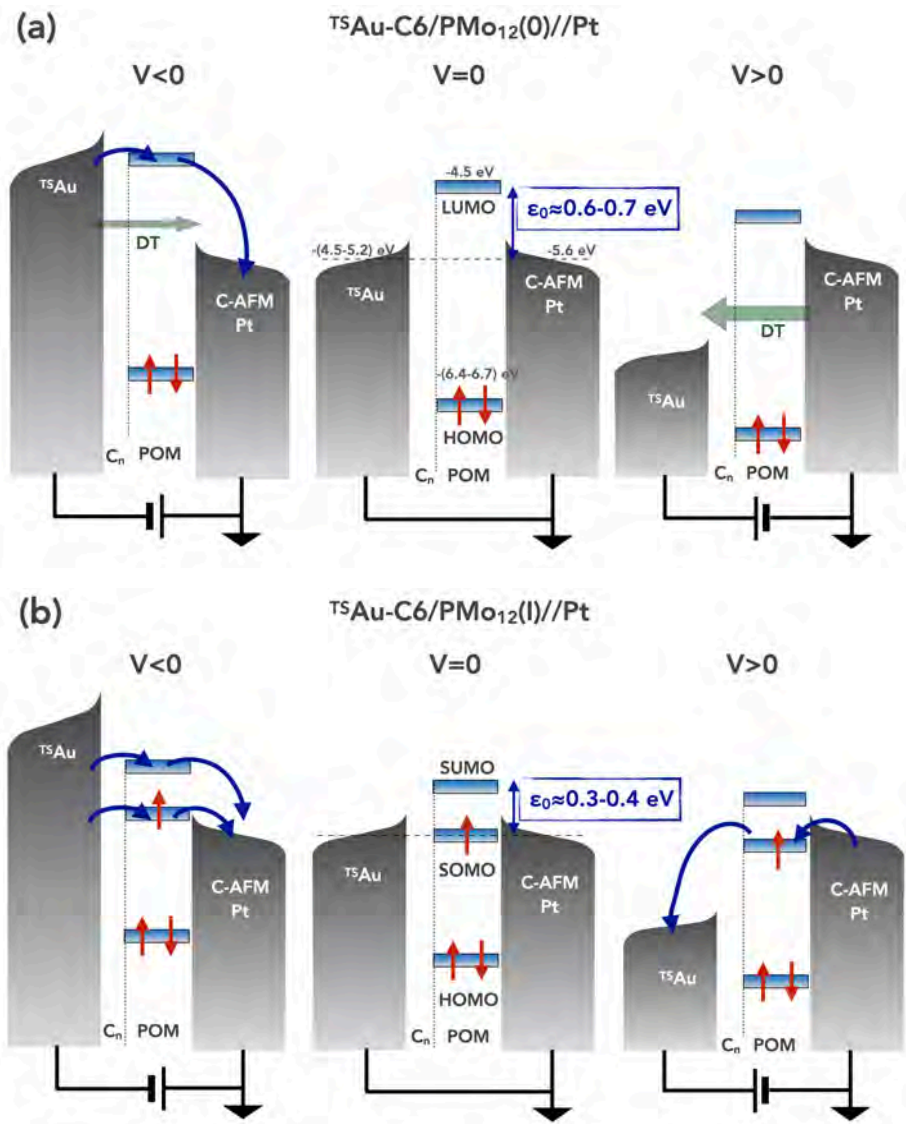

*Figure 6.* Hypothesized schemes of the energy diagrams : (a) $^{TS}$Au-C6/PMo$_{12}$(0)//Pt and (b) $^{TS}$Au-C6/PMo$_{12}$(I)//Pt junctions at a negative, null and positive bias applied on the $^{TS}$Au substrate (the C-AFM tip grounded). The blue arrows indicate the MO-mediated electron transport and the green arrows show the direct tunneling (the wider the arrow, the larger the tunneling current). The MO energies (vs. vacuum energy) are given as reported from calculations (see text) and the $\varepsilon_0$ values are taken from our experiments (Figs. 2 and 4, Table 2).

The ML and clustering methods allow us to refine the analysis of the large dispersion of the I-V dataset. For the $^{TS}$Au-C6/PMo$_{12}$(0)//Pt, the two clusters with the highest current level (#2 and #3, 0.1 - 1 nA at +/- 1.5V, Fig. 5b) have the same LUMO energy (≈ 0.6 eV) in agreement with the average value deduced from statistics on the full dataset (Fig. 4). They only differ by the electronic coupling to the electrodes (a factor about 4.5), which likely reflects fluctuations of the C-AFM tip contact on the SAM. The two other clusters at low current (#4 and #5, ≈ 10 pA at +/- 1.5 V, Fig. 5b) have more dispersed values of the LUMO but they likely include noisy I-V traces near the sensitivity limit of the current-voltage preamplifier (Fig. S16). We note, however, that the cluster 3 displays an asymmetric shape of the I-V in contrast to the I-Vs of the other clusters and the global mean Ī-V (Fig. 2, Fig. S16, Table 3). For the $^{TS}$Au-C6/PMo$_{12}$(I)//Pt, we can distinguish two series of two clusters each. In the first series, the clusters 2 and 3 with the highest current level (≈ 10 nA at +/- 1.5 V, Fig. 5d) are characterized by a MO level at 0.38-0.43 eV with the largest electronic coupling to the electrodes (0.35-1.1 meV), and, in the second series, the clusters 4 and 5 show a slightly lower energy level (≈ 0.3 eV), but a worse electrode coupling (0.07-0.1 meV) leading to the lowest current level (≈ 10-100 pA et +/- 1.5 V, Fig. 5d). This feature may be due to different configurations of the PMo$_{12}$(I) between the C6 SAM and the C-AFM tip, as well as, fluctuations of the C-AFM tip contact (not easily distinguishable at this experimental level). The main difference compared to the PMo$_{12}$(0) case is that all the clusters (Fig. S21) display asymmetric I-Vs (as for the global mean Ī-V in Fig. 2) with the largest asymmetry for clusters 4 and 5. Thus this asymmetric feature in these $^{TS}$Au-C6/PMo$_{12}$//Pt junctions can be viewed as a finger print of the reduced PMo$_{12}$(I) and the asymmetric cluster 3 for the PMo$_{12}$(0) device could be ascribed to the presence of a small fraction of reduced Mo in that case as observed from XPS (see Fig. S5 in the SI).

## Conclusion.

We investigated the electronic properties of switchable redox polyoxometalates (phosphomolybdate, $PMo_{12}$) by a combination of i) electron transport measurements at the nanoscale (C-AFM on self-assembled monolayers), ii) analytical models statistically applied on large current-voltage datasets and iii) unsupervised machine learning and clustering algorithms. The main results are summarized as follows:

1. We demonstrate a reversible redox switching triggered *in situ* by UV photoreduction.

2. The one-electron reduced $PMo_{12}$ (open-shell state) is stable in the molecular junctions in air and at room temperature, the spontaneous return to its oxidized state laps for several hours to days.

3. The reduced $PMo_{12}$ molecular junctions are characterized by an increase of the conductance (a factor ∼ 10) and asymmetric current-voltage curves.

4. The electron transport in the pristine $PMo_{12}$ junctions is controlled by the LUMO located at ∼ 0.6-0.7 eV above the Fermi energy of the electrodes, with 25% of the junction dataset characterized by a better electronic coupling to the electrodes.

5. The electron transport in the reduced $PMo_{12}$ junctions and the asymmetric current-voltage behavior is ascribed to a combined electron transmission through energetically closed (0.3-0.4 eV) SOMO and SUMO levels near resonance with the Fermi energy of the electrodes. This latter point calls for detailed theoretical studies.

## Materials and methods

### *Synthesis and sample fabrication.*

*Molecule synthesis.* The $PMo_{12}(0)$ and $PMo_{12}(I)$ were prepared as previously reported by some of us,[16] and the characterizations ($^{31}P$ NMR, cyclic voltammetry, solution UV-vis spectroscopy, XPS spectra on powder) are given in the Supporting Information (section 1).

*Bottom metal electrode fabrication.* Template stripped gold ($^{TS}$Au) substrates were prepared according to the method previously reported.[17, 74, 75] In brief, a 300–500 nm thick Au film is evaporated on a very flat silicon wafer covered by its native $SiO_2$ and then transferred to a glued clean glass piece which is mechanically stripped with the Au film attached on the glass piece, letting exposed a very flat (RMS roughness of 0.4 nm, the same as the starting $SiO_2$ surface used as the template).

*Self-assembled monolayers.* The SAMs on $^{TS}$Au were fabricated following a protocol developed and optimized in a previous work for the electrostatic immobilization of POMs on amine-terminated SAMs.[11] In brief, we first dipped the freshly prepared metal substrate in a solution of 6-aminohexane-1-thiol hydrochloride (HS-$(CH_2)_6$-$NH_2$) at a concentration of $10^{-3}$ M in ethanol overnight in the dark (see details in section 2 in the Supporting Information). Then the samples were dipped in a solution of POMs at a concentration of $10^{-3}$ M in acetonitrile for several hours (we checked that the thickness of the POM layer was independent of the immersion time when > 1h).

**Spectroscopic ellipsometry.**

The thickness of the SAMs was measured by spectroscopic ellipsometry (UVISEL ellipsometer (HORIBA), section 3 in the Supporting Information).

**C-AFM in ambient conditions.** We measured the electron transport properties at the nanoscale by C-AFM (ICON, Bruker) at room temperature using a tip probe in platinum (with loading force of ca. 6-8 nN). The voltage was applied on the substrate, the tip is grounded via the input of the current-voltage preamplifier. We used a "blind" mode to measure the current-voltage (I-V) curves and the current histograms: a square grid of 10×10 points was defined with a pitch of 50 to 100 nm. At each point, the I-V curve (back and forth) is acquired leading to the measurements of 200 traces per grid. This process was repeated 3 times at different places (randomly chosen) on the sample, and up to 600 I-V traces were used to construct the current-voltage histograms (section 4 in the Supporting Information).

***Photoreduction.***

A UV lamp (Analytik Jena) was used for the UV light irradiation at 302 nm for the CAFM measurements (section 5 in the Supporting Information).

## Associated content

The Supporting Information is available free of charge at xxxxxx.

- Details on synthesis, RMN, cyclic voltammetry, UV-vis spectroscopy, XPS, fabrication of electrodes and self-assembled monolayers, ellipsometry, C-AFM data analysis and fit protocols, UV illumination setup, I-V curves of the redox cycles, machine learning and clustering details on the validation and analysis of the different clusters of I-V traces, data for reference samples (without POMs).

***Author Contributions***

C.H. and D.G. fabricated the SAMs and performed the physicochemical characterizations (UV-vis, XPS), F.V. and A.P. synthesized and characterized the POMs. C.H. did all the CAFM measurements, C.H. and D.V. analyzed the data. The machine learning analysis was conducted in the framework of a collaboration with M.C., C.H. under the supervision of M.L.P. performed the machine learning analysis. A.P. and D.V. conceived and supervised the project. This work is part of the PhD thesis of C.H., S.L. and D.V. supervised the thesis. The manuscript was written by D.V. with the contributions of all the authors. All authors have given approval of the final version of the manuscript.

***Note***

The authors declare no competing financial interest.

**Acknowledgements.**

We acknowledge support of the CNRS, project "neuroPOM", under a grant of the 80PRIME program. We acknowledge Xavier Wallart (IEMN-CNRS) for his help with the XPS measurements. K. Trinh (IPCM-CNRS) is warmly acknowledged for the synthesis of PMo$_{12}$(0) and PMo$_{12}$(I).

# Redox-Controlled conductance of polyoxometalate molecular junctions.

Cécile Huez,[1] David Guérin,[1] Stéphane Lenfant,[1] Florence Volatron,[2] Michel Calame,[3,4] Mickael L. Perrin,[3,5] Anna Proust[2] and Dominique Vuillaume.[1]

1) Institute for Electronics Microelectronics and Nanotechnology (IEMN), CNRS, University of Lille, Av. Poincaré, Villeneuve d'Ascq, France.
2) Institut Parisien de Chimie Moléculaire (IPCM), CNRS, Sorbonne Université, 4 Place Jussieu, F-75005 Paris, France.
3) EMPA, Transport at the Nanoscale Laboratory, 8600 Dübendorf, Switzerland.
4) Dept. of Physics and Swiss Nanoscience Institute, University of Basel, Klingelbergstrasse 82, 4056 Basel, Switzerland.
5) Department of Information Technology and Electrical Engineering, ETH Zurich, 8092 Zurich, Switzerland.

**SUPPORTING INFORMATION**

**Section 1. Synthesis and characterization of $(TBA)_3[PMo^{VI}_{12}O_{40}]$ ($PMo_{12}(0)$) and $(TBA)_4[PMo^{VI}_{11}Mo^{V}_{1}O_{40}]$ ($PMo_{12}(I)$).**

$PMo_{12}(0)$ and $PMo_{12}(I)$ have been synthesized according to previously published procedures.[1, 2] The purity of the compounds was confirmed by IR and $^{31}P$ NMR spectroscopies.

**Synthesis of $PMo_{12}(0)$**: 60 mL of a 1M solution of sodium molybdate dihydrate $Na_2[MoO_4].2H_2O$ was added to 9 mL of nitric acid $HNO_3$ and 50 mL of 1,4-dioxane. Under stirring, 5 mL of a 1 M solution of orthophosphoric acid $H_3PO_4$ and 5g of tetrabutylammonium bromide $NBu_4Br$ (TBABr) are added. After filtration, the yellow heavy solid is immersed in 50 mL of boiling water and stirred, filtered again and washed with 50 mL water, 100 mL ethanol and diethyl ether until obtaining a yellow powder. It is finally recrystallized in hot acetone: 30 mL of hot acetone are required to recrystallize 1 g of powder. After three days in the refrigerator, the mixture is filtered and yellow crystals are collected, dried under vacuum several days at 60°C. Note that during the experiment, the POM was handled with glass spatula/material to avoid its reduction. IR (KBr, cm$^{-1}$): 2962 (m), 2933 (m), 2873 (m), 1473 (m), 1381 (w), 1063 (s), 967 (shoulder), 956 (vs), 880 (s), 806 (vs), 739 (w), 619 (w), 504 (m), 465 (w), 387 (s), 342 (m). $^{31}P$ NMR (400 MHz, $CD_3CN$, ppm): δ = -3.29 (s).

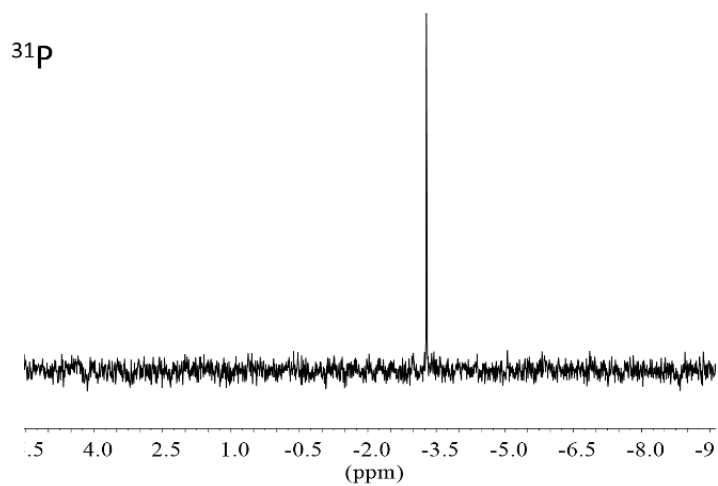

**Figure S1**. $^{31}P$ NMR spectrum of $PMo_{12}(0)$ recorded in $CD_3CN$.

**Synthesis of $PMo_{12}(I)$**: 100 mg of $(TBA)_3[PMo_{12}O_{40}]$ are dissolved in a minimum volume of dry acetonitrile (~6 mL) in a dry Schlenk tube containing a magnetic stir bar, under Argon. Under stirring, a few drops of phenyllithium are added to the $PMo_{12}(0)$ solution (color change from yellow to green). The reaction is followed by recording $^{31}$P NMR spectra in which the initial singlet at -3.29 ppm slowly disappears. Drops of phenyllithium is added until the appearance of a signal at 0.49 ppm corresponding to PMo$_{12}$(I), and the solution displays a blue color. 15 mg of NBu$_4$Br are added to the solution followed by the addition of ~15 mL of diethylether, leading to the formation of a blue precipitate. The suspension is filtered on a cellulose membrane. The blue solid is subsequently washed with 10 mL of tetrahydrofuran and 10 mL of methanol. It is finally dried under vacuum, in the dark. $^{31}$P NMR (162 MHz, CD$_3$CN, ppm): δ = 0.49 (s).

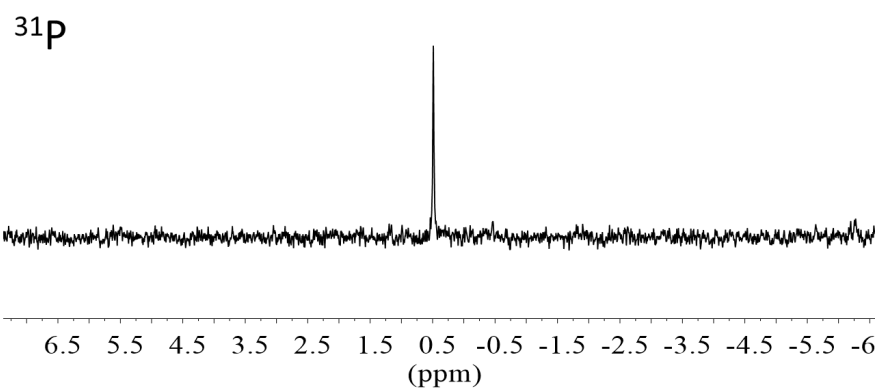

**Figure S2**. $^{31}$P NMR spectrum of PMo$_{12}$(I) recorded in CD$_3$CN.

The electrochemical behavior of a 1 mM solution of (TBA)$_3$[PMo$^{VI}_{12}$O$_{40}$] in 0.1 M TBAPF$_6$ in CH$_3$CN was checked by cyclic voltammetry. A standard three electrode cell was used, which consisted of a working vitreous carbon electrode, an auxiliary platinum electrode and an aqueous saturated calomel electrode (SCE) equipped with a double junction to allow its use in an organic solvent. In those conditions the redox potential can be equally given versus SCE (0.308 V versus NHE) or recalculated versus the Fc$^+$/Fc couple (0.690 V versus NHE).[3] The cyclic voltammogram is displayed below (Fig. S3). It features three reversible monoelectronic processes with midpoint potentials E$_{1/2}$=0.5(E$_{pa}$-E$_{pc}$) (E$_{pa}$: anodic peak potential; E$_{pc}$ cathodic peak potential) at +0.142, -0.272 and -0.991 V/SCE (respectively -0.24, -0.654 and -1.373 V versus Fc$^+$/Fc). The LUMO energy position is calculated by $E_{LUMO}=-(E_{1/2red}+E_{ref/ESH})-4.44$, with $E_{ref/ESH}=0.308$ eV for the saturated calomel electrode (SCE) in CH$_3$CN.3 With $E_{1/2red}=0.142$ V/SCE for the one-electron reduction, we get $E_{LUMO}$ = -4.89 eV with respect to the vacuum level.

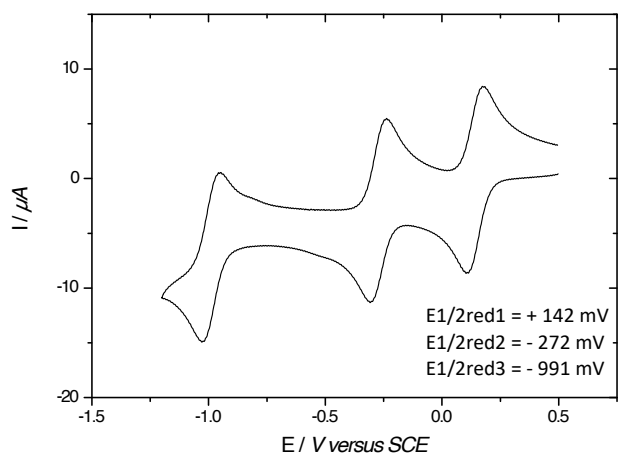

**Figure S3.** *Cyclic voltammetry showing reduction of the PMo$_{12}$ up to 3 electrons.*

The redox state was also characterized by UV-vis spectroscopy of POMs in solution (≈ µM in CH$_2$Cl$_2$). UV-Vis absorption spectra (Fig. S4) were recorded on a Lambda 800 Perkin-Elmer spectrometer. For the reduced PMo$_{12}$(I), we clearly observe a shift of the LMCT (ligand-to-metal charge transfer) band (309 → 315 nm) and the appearance of the IVCT (intervalence charge transfer) band between Mo(V) and Mo(VI) at around 750 nm.

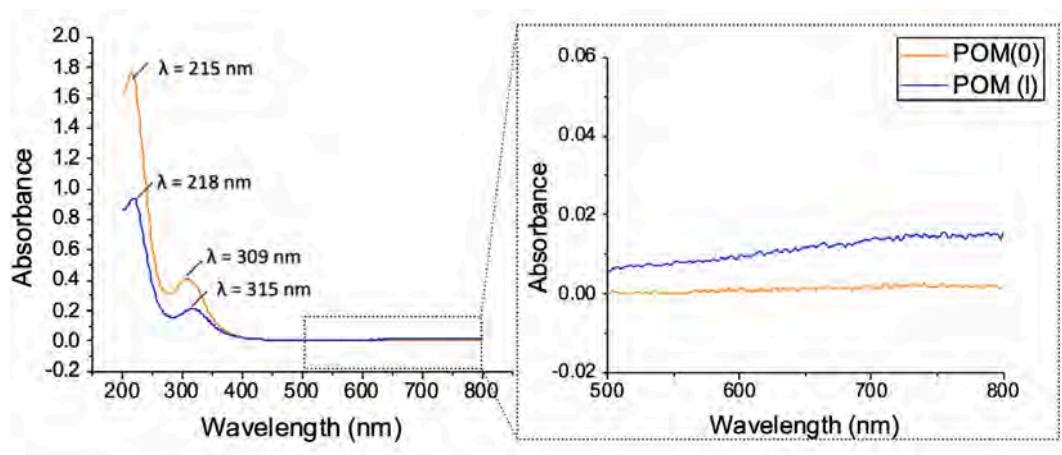

**Figure S4.** UV-vis absorbance spectra of PMo$_{12}$(0) and PMo$_{12}$(I) in solution.

High resolution XPS spectra were recorded with a monochromatic Al$_{K\alpha}$ X-ray source (hυ = 1486.6 eV), a detection angle of 45° as referenced to the sample surface, an analyzer entrance slit width of 400 μm and with an analyzer pass energy of 12 eV. In these conditions, the overall resolution as measured from the full-width half-maximum (FWHM) of the Ag 3d$_{5/2}$ line is 0.55 eV. Background was subtracted by the Shirley method.[4] The peaks were decomposed using Voigt functions and a least squares minimization procedure. Binding energies (BE) were referenced to the C 1s BE, set at 284.8 eV. The XPS measurements were done on powder of PMo$_{12}$(0) and PMo$_{12}$(I) deposited on Si/SiO$_2$ functionalized with APTES (aminopropyltriethoxysilane) that give a better uniform deposition of the powder than on Au functionalized with aminoalkylthiol. The figure S5 shows the Mo 3d spectra. The energy splitting between the 3d 3/2 and 3d 5/2 peaks is fixed to 3.15 eV with an amplitude ratio of 0.67.[5] From the peak areas, we calculate the Mo(VI)/Mo(V) ratios of 12.1 for PMo$_{12}$(0) and 3.9 for PMo$_{12}$(I).

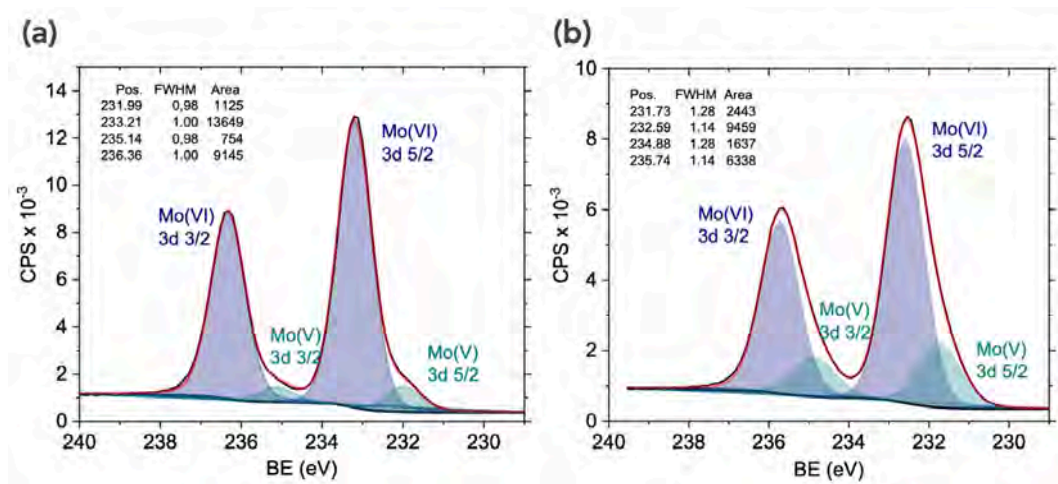

*Figure S5*. XPS spectra (Mo 3d) of (a) PMo$_{12}$(0) and (b) PMo$_{12}$(I).

**Section 2. Electrodes and SAMs fabrication.**

*Electrodes.*

Ultraflat template-stripped gold surfaces ($^{TS}$Au), with rms roughness of ~0.4 nm were prepared according to the method already reported.[6-8] In brief, a 300–500 nm thick Au film was evaporated on a very flat silicon wafer covered by its native SiO$_2$ (rms roughness of ~0.4 nm), which was previously carefully cleaned by piranha solution (30 min in 7:3 H$_2$SO$_4$/H$_2$O$_2$ (v/v); **Caution**: Piranha solution is a strong oxidizer and reacts exothermically with organics), rinsed with deionized (DI) water, and dried under a stream of nitrogen. Clean 10x10 mm pieces of glass slide (ultrasonicated in acetone for 5 min, ultrasonicated in 2-propanol for 5 min, and UV irradiated in ozone for 10 min) were glued on the evaporated Au film (UV-polymerizable glue, NOA61 from Epotecny), then mechanically peeled off providing the $^{TS}$Au film attached on the glass side (Au film is cut with a razor blade around the glass piece).

*Self-assembled monolayers.*

The self-assembled monolayers (SAMs) of 6-aminohexane-1-thiol (HS-(CH$_2$)$_6$-NH$_2$) were prepared following a protocol optimized and described in a previous work for the electrostatic immobilization of POMs on amine-terminated SAMs.[9] The freshly prepared $^{TS}$Au substrates were dipped in a solution of 6-aminohexane-1-thiol hydrochloride (Sigma-Aldrich) at a concentration of $10^{-3}$ M in ethanol overnight in the dark. The samples were rinsed in ethanol for 5 min and then ultrasonically cleaned 5 min in deionized (DI) water. These SAMs were treated by a PBS (phosphate-buffered saline, pH=7.4) solution for 2 hours, followed by ultra-sonication in DI water for 5 minutes. The substrates were finally washed with ethanol and dried under nitrogen flow. It was found that the PBS treatment removes the formation of aggregates on the aminoalkylthiol SAMs as well as avoids clustering of POMs during the electrostatic deposition, likely because this treatment optimizes the ratio of $NH_3^+/NH_2$ on the surface.[9] The electrostatic deposition of $PMo_{12}(0)$ and $PMo_{12}(I)$ was done by immersion of these SAMs in a solution of $PMo_{12}$ at a concentration of $10^{-3}$ M in acetonitrile for one to few hours. We checked by ellipsometry that the thickness of the POM layer was independent of the immersion time when the immersion time is longer than 1h. It is not possible to distinguish by XPS the N atoms from TBA and from the protonated amine-terminated SAM. We crudely estimated the ratio $TBA^+/NH_3^+$ ensuring the global electrical neutrality as follows. A perfectly, closely packed, SAM of alkyl chains in Au surface has a maximum density of $4 \times 10^{14}$ chain/cm$^2$ (or one alkyl chain per 25 Å$^2$)[10] and assuming a protonation at 30%, we estimated a density of $1.2 \times 10^{14}$ $NH_3^+$/cm$^2$. The most dense layer of POMs (sphere with a diameter of 1 nm) is a centered hexagonal geometry with a concentration of $\sim 1.15 \times 10^{14}$ POM/cm$^2$ (one POM per 86 Å$^2$), thus about one $NH_3^+$ per POM. In this ideal case, the neutrality requires 1 $NH_3^+$ and 2 $TBA^+$ counterions for $PMo_{12}(0)$ or 3 $TBA^+$ counterions for $PMo_{12}(I)$, this $TBA^+/NH_3^+$ ratio could be lower for a less dense POM layer.

**Section 3. Spectroscopic ellipsometry.**

We recorded spectroscopic ellipsometry data (on *ca.* 1 cm$^2$ samples) in the visible range using a UVISEL (Horiba Jobin Yvon) spectroscopic ellipsometer equipped with DeltaPsi 2 data analysis software. The system acquired a spectrum ranging from 2 to 4.5 eV (corresponding to 300–750 nm) with intervals of 0.1 eV (or 15 nm). The data were taken at an angle of incidence of 70°, and the compensator was set at 45°. We fit the data by a regression analysis to a film-on-substrate model as described by their thickness and their complex refractive indexes. First, a background for the substrate before monolayer deposition was recorded. We acquired three reference spectra at three different places of the surface spaced of few mm. Secondly, after the monolayer deposition, we acquired once again three spectra at three different places of the surface and we used a 2-layer model (substrate/SAM) to fit the measured data and to determine the SAM thickness. We employed the previously measured optical properties of the substrate (background), and we fixed the refractive index of the organic monolayer at 1.50.[10] We note that a change from 1.50 to 1.55 would result in less than a 1 Å error for a thickness less than 30 Å. The three spectra measured on the sample were fitted separately using each of the three reference spectra, giving nine values for the SAM thickness. We calculated the mean value from this nine thickness values and the thickness incertitude corresponding to the standard deviation. Overall, we estimated the accuracy of the SAM thickness measurements at ± 2 Å.[11]

**Section 4. AFM measurements.**

*TM-AFM.*

Topographic images were acquired in tapping mode (TM) on an ICON (Bruker) microscope using a silicon tip (42 N/m spring constant, resonance frequency 320 kHz) at room temperature and in ambient condition. The AFM images were treated with the Gwyddion software.[12]

*C-AFM.*

Current–voltage characteristics were measured by conductive atomic force microscopy (Icon, Bruker), using Pt coated tip (RMN-12PT400B from Bruker, 0.3 N/m spring constant). To form the molecular junction, the conductive tip was located at a stationary contact point on the SAM surface at controlled loading force (~ 6-8 nN). The voltage was applied on the substrate, the tip is grounded via the input of the current-voltage preamplifier. The C-AFM tip is located at different places on the sample (typically on an array - 10x10 grid - of stationary contact points spaced of 50-100 nm), at a fixed loading force and the I–V characteristics were acquired directly by varying voltage for each contact point. The I-V characteristics were not averaged between successive measurements and typically up to 600 I-V measurements were acquired on each sample.

*Loading force and C-AFM tip contact area.*

The load force was set at ≈6-8 nN for all the I-V measurements, a lower value leading to too many contact instabilities during the I-V measurements. Albeit larger than the usual load force (2-5 nN) used for CAFM on SAMs, this value is below the limit of about 60-70 nN at which the SAMs start to suffer from severe degradations. For example, a detailed study (Ref. 13) showed a limited strain-induced deformation of the monolayer (≤ 0.3 nm) at this used load force. The same conclusion was confirmed by our own study comparing mechanical and electrical properties of alkylthiol SAMs on flat Au surfaces and tiny Au nanodots.[14]

*Data analysis.*

Before to construct the current histograms and analyze the I-V curves with the one energy-level model and the TVS method, the raw set of I-V data is scanned and some I-V curves were discarded from the analysis:

- At high current, the I-V traces that reached the saturating current during the voltage scan (the compliance level of the trans-impedance amplifier, typically $5\times10^{-9}$ A in Figs. S6, depending on the gain of the amplifier) and/or I-V traces displaying large and abrupt steps during the scan (contact instabilities).

- At low current, the I-V traces that reached the sensitivity limit (almost flat I-V traces and noisy I-Vs) and displayed random staircase behavior (due to the sensitivity limit - typically 0.1-1 pA depending on the used gain of the trans-impedance amplifier and the resolution of the ADC (analog-digital converter), Fig. S6.

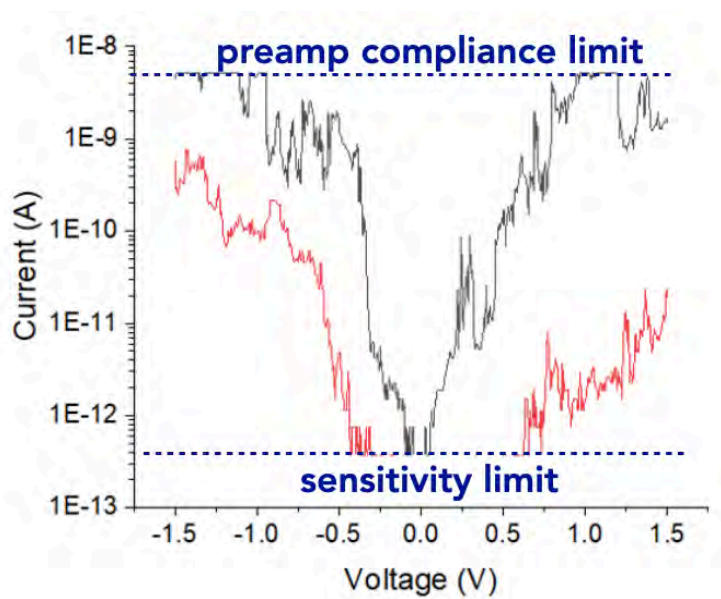

*Figure S6*. Typical examples of I-V curves discarded from the data analysis.

***Fit of the single energy level (SEL) model.***

All the I-V traces in Fig. 2 (main text) were fitted individually with the single energy-level (SEL) model (Eq. 1, main text) with 3 fit parameters: $\varepsilon_{0\text{-SEL}}$ the energy position (with respect to the Fermi energy of electrodes) of the molecular orbital involved in the electron transport, $\Gamma_1$ and $\Gamma_2$ the electronic coupling energy between the molecules and the two electrodes. The fits (Figs. S7a and S7b) were done with the routine included in ORIGIN software, using the method of least squares and the Levenberg Marquardt iteration algorithm.

The SEL model is a low temperature approximation albeit it can be used at room temperature for voltages below the resonant transport conditions[15, 16] since the temperature broadening of the Fermi function is not taken into account. Moreover, a possible voltage dependence of $\varepsilon_{0\text{-SEL}}$ is also neglected.[17] It is known that the value of $\varepsilon_{0\text{-SEL}}$ given by the fit of the SEL model depends on the voltage window used for the fit.[15-17] This feature is confirmed (Fig. S8) showing that unreliable values are obtained with a too low voltage range (i.e. the SEL model is not reliable in the linear regime of the I-V curves) and not applicable when the voltage is high enough to bring the electrode Fermi energy close to molecular orbital (near resonant transport), here for a voltage window larger than -1.2/1.2V V where the fits are bad and the values of $\varepsilon_{0\text{-SEL}}$ collapse. For voltage windows below -1V/1V we clearly see the lowering of $\varepsilon_{0\text{-SEL}}$ by about 0.15-0.2 eV after reduction of the POMs. For comparison, the same mean $\bar{I}$-V curves are also analyzed by TVS (Fig. S7). We obtain a good agreement with the SEL model limiting the fit in the voltage window -1V/1V. For these reasons we limited the fits to a voltage window -1 V to 1 V to analyze the complete datasets shown in Fig. 2 (main text). To construct the histograms of the values of $\varepsilon_{0\text{-SEL}}$, $\Gamma_1$ and $\Gamma_2$ (Fig. 4 in the main text), we discarded the cases for which the fits were not converging of not accurate enough (i.e. R-squared < 0.95). In the case of the reduced POM(I), the SEL model does not fit the data whatever the voltage window considered (Figs. S7 and S8), likely because the OM involved in the transport is too close to the Fermi energy (≈0.3 eV determined by TVS, Fig. S7), a situation where the model is not valid. Thus only the TVS method was used to analyze the I-V dataset in this latter case.

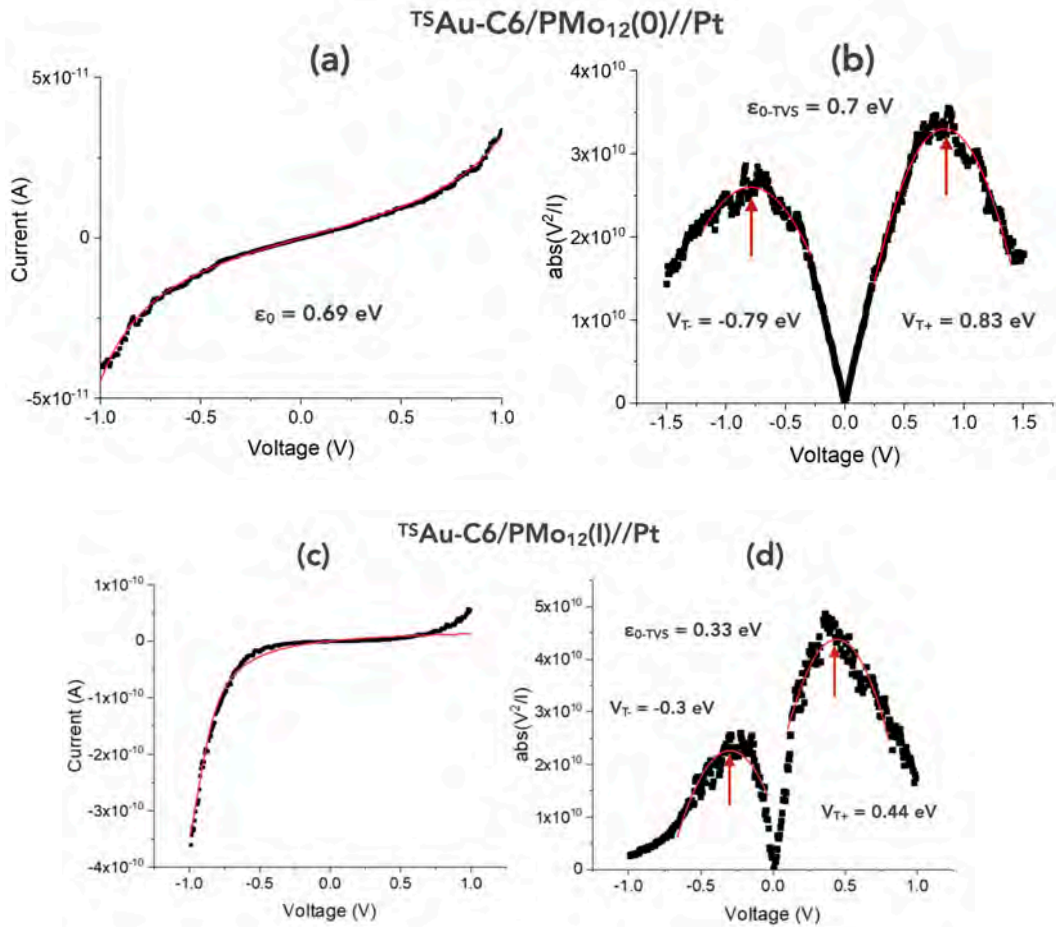

*Figure S7.* One energy-level model fits on the mean current-voltage curves within the bias voltage range -1V/1V (solid red line) for (a) the $^{TS}$Au-C6/PMo$_{12}$(0)//Pt and (c) the $^{TS}$Au-C6/PMo$_{12}$(I)//Pt junctions. Typical TVS plots (|V²/I|) vs. V for (b) the $^{TS}$Au-C6/PMo$_{12}$(0)//Pt and (d) the $^{TS}$Au-C6/PMo$_{12}$(I)//Pt junctions. The thresholds $V_{T+}$ and $V_{T-}$ are indicated by the vertical lines (with values) - max of a 2nd order polynomial function fitted around the max of the bell-shaped curves (to cope with noisy curves).

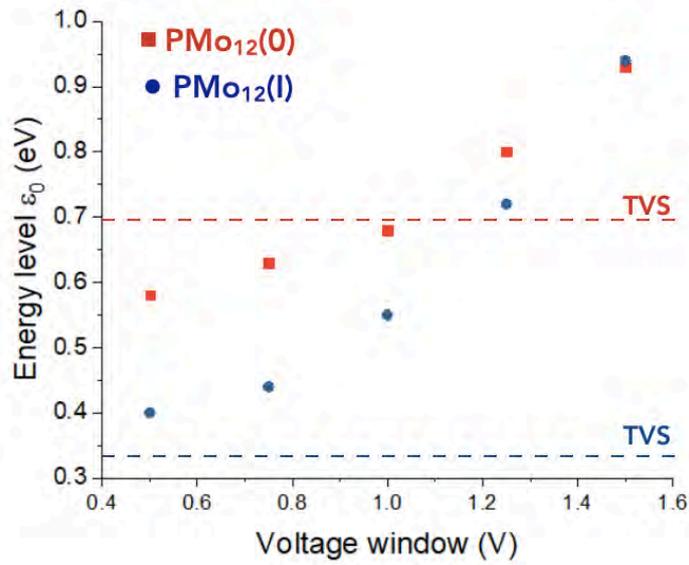

***Figure S8.*** *Values of $\varepsilon_{0\text{-}SEL}$ obtained with the SEL model fitted on the mean $\bar{I}$-V curves for the $^{TS}$Au-C6/PMo$_{12}$(0)//Pt and $^{TS}$Au-C6/PMo$_{12}$(I)//Pt with increasing voltage windows (-0.5/0.5 V to -1.5/1.5 V) for the fits. The dashed lines indicate the value obtained by the TVS method (Fig. S7).*

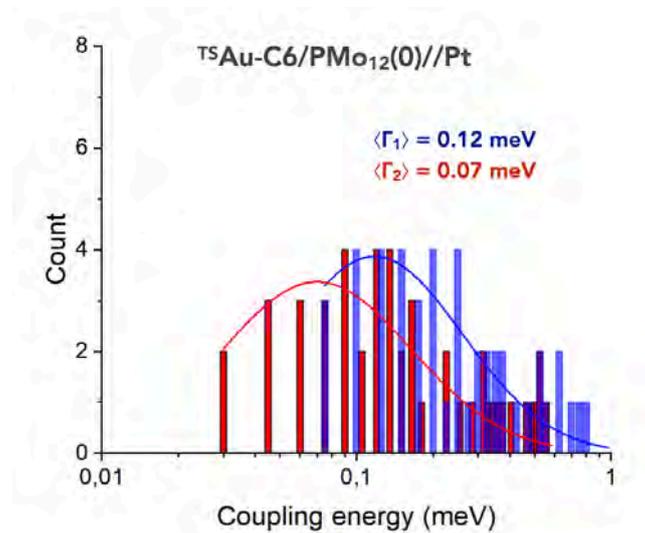

***Figure S9.*** *Distribution of the electrode coupling parameters $\Gamma_1$ and $\Gamma_2$ (SEL model) of the I-V data for the $^{TS}$Au/PMo$_{12}$(0)//Pt junctions (Fig. 2, main text) and log-normal fits of these distributions.*

**Section 5. Illumination setup.**

We used a UV lamp (Analytic Jena) for UV light irradiation. This lamp has a wavelength centered at 302 nm (close to the absorbance peak, see Fig. S4) with a power of 0.5 mW/cm$^2$. The lamp was brought close (*ca.* few centimeters) to the sample in the CAFM setup.

**Section 6. Redox cycles.**

Figure S10 shows the 2D histograms of the same $^{TS}$Au/PMo$_{12}$(0)//Pt junction (pristine) sequentially irradiated by UV light and let to relax in air at RT or under a moderate heating (hot plate, 80°C), the corresponding current histograms at -1V fitted by a log-normal distribution. The log-mean current ± the log-standard deviation is plotted in Fig. 3 (main text) versus the three sequences of photoreduction/relaxation. We note a degradation (lower currents) for the third cycle (data #6-8 in Figs. 3 and S10), albeit with a similar effect of the irradiation on the conductance. This degradation (lower currents than for the pristine sample) is observed whatever the reoxidation step (RT or heating at 80°C), while during the second cycle, the step at 80°C (data #5) returned the sample to almost the same current level as the pristine one. Thus the heating at 80°C is not specifically responsible to the sample degradation. Alkylthiol SAMs on Au are thermally stable up to 100-150°C,[18] and up to ~200°C for PMo$_{12}$.[19] Moreover, a thermal degradation (*e.g.* molecule desorption from the surface or thermal decomposition of the molecules) would have induced an increase of the current (less molecules in the SAMs and consequently a thinnest SAM, or even a direct contact between the C-AFM tip and the underlying Au substrate). This global decrease of the current (both for the reduced and oxidized samples) during the third cycle might be due to a drift of the loading force (e.g. a small decrease of the loading force), a pollution of the C-AFM tip or a pollution of the sample during the long duration of these measurements. A control measurement of the SAM thickness (ellipsometry) after the three reduction/oxidation steps showed that thickness has increased by ~ 1 nm compared to the one of the pristine sample.

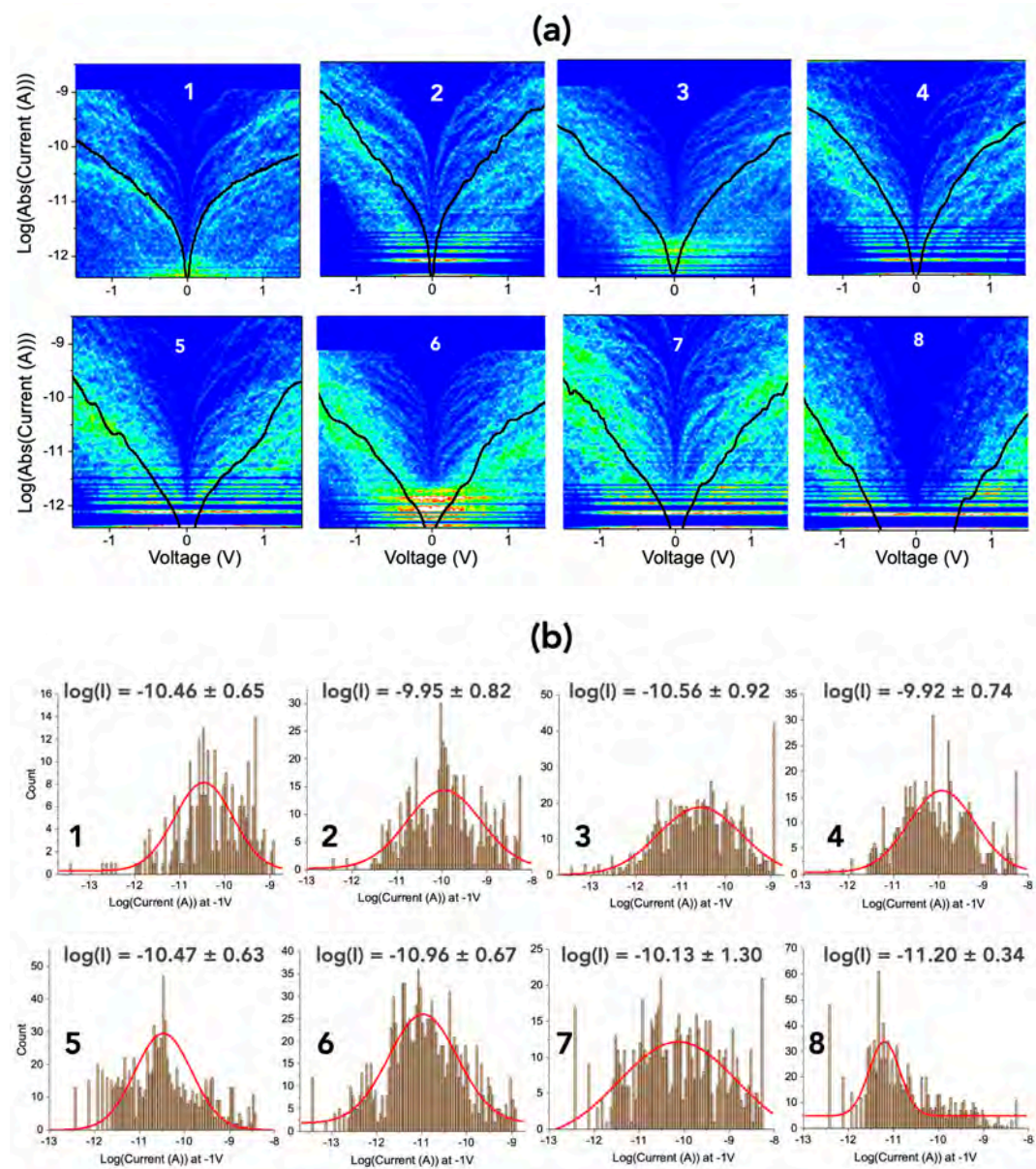

*Figure S10*. *(a) 2D histograms of the I-V curves and (b) histograms of the current at -1V fitted by a log-normal distribution (the log-mean current and log-standard*

*deviation are given in the panels) for: 1) pristine, 2) UV 5.5h, 3) RT 14.5h, 4) UV 4h, 5) RT 24h, 6) UV 3.5h, 7) 80°C 2h.*

In addition, the PMo$_{12}$(0) molecules were drop cast on a glass substrate and the film was UV irradiated in the same condition as the SAMs. The film turned from yellow-like to green-like (Fig. S11) indicating a partial reduction of the film (green = PMo$_{12}$(0) yellow + PMo$_{12}$(I) blue). It returns yellow-like after exposition in air, indicating the reversibility of the redox switching.

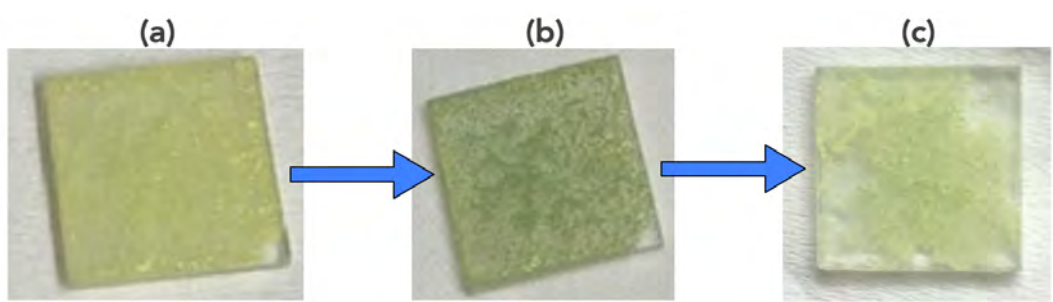

***Figure S11**. Pictures of a drop cast films (a) before, (b) after UV irradiation for 6 h and (c) after few days exposed at air.*

### Section 7. Machine learning and clustering.

*Rationalized choice of the number of clusters.*

To fix the optimized number of clusters, we analyzed the same dataset with various number of clusters from 2 to 6. Fig. S12 shows the obtained mean Ī-Vs for the dataset of $^{TS}$Au-C6/PMo$_{12}$(0)//Pt junctions for 4, 5 and 6 clusters (the clusters are labeled cN/M, with N the cluster number by decreasing order of current amplitude and M the total number of clusters). In all cases, the cluster 1 corresponds to I-Vs saturating (current-voltage preamplifier compliance) during the measurements and this cluster is not considered further in the analysis. The mean Ī-V curve of each cluster was analyzed with the SEL model and TVS method and the obtained MO energy levels are given in Table S1. Clearly, the solution with 4 clusters (and less, not shown) is not satisfactory because the cluster c3/4 can be decomposed (see the feature spaces in panels (a) and (b) in Fig. S12) in two clusters (c3/5 and c4/5) with significant differences: i) the mean $\bar{I}$-V curve of c3/5 displays a negative asymmetry, while the cluster c4/5 shows an almost symmetric mean $\bar{I}$-V curve (panel (e) in Fig. S12); (ii) the $\varepsilon_0$ values (SEL and TVS) are different for c3/5 than c4/5 (see Table S1 and Table 3 in the main text), as well as the electronic coupling to the electrodes (Γ values, see Table 3 in the main text), while the values for the c3/4 clusters are intermediate between those of the c3/5 and c4/5 clusters. Thus the analysis with 5 clusters is more pertinent. Extending to 6 clusters splits the cluster c2/5 in two (c2/6 and c3/6, see feature spaces in panels (b) and (c) in Fig. S12, but the deduced $\varepsilon_0$ values (SEL and TVS) are similar (see Table S1), while the other clusters are not modified by extending to 6 clusters (Figs. S12 and Table S1: c3/5 ≡ c4/6, c4/5 ≡ c5/6 and c5/5 ≡ c6/6). Thus using 6 clusters does not add more pertinent information and we conclude that the analyze with 5 clusters is the optimized approach.

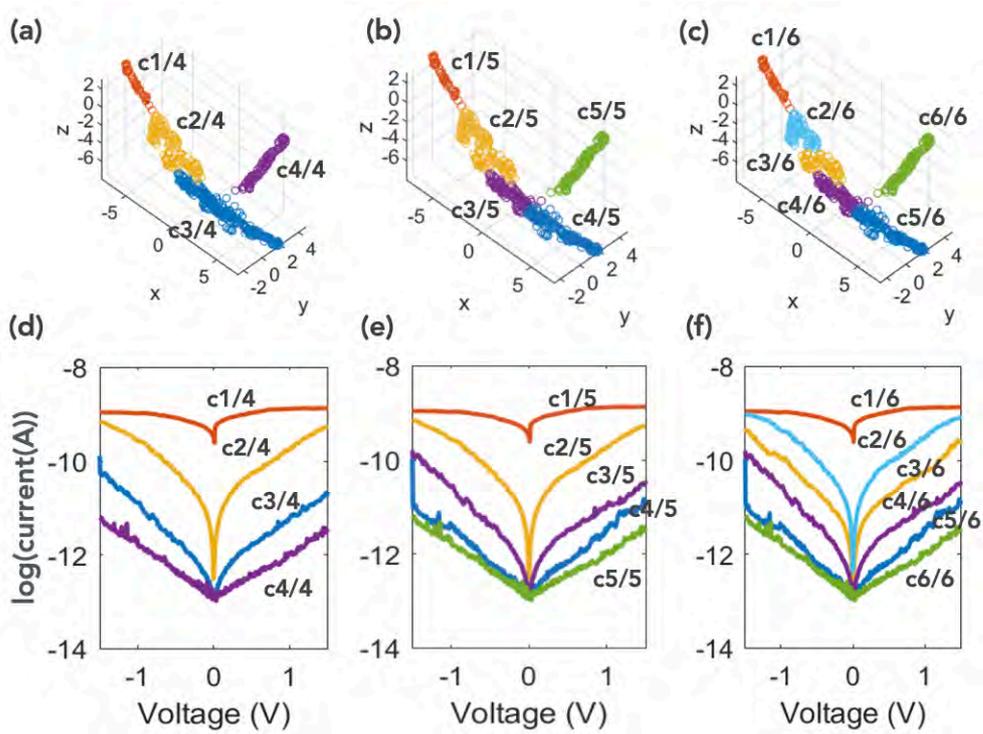

**Figure S12.** $^{TS}$Au-C6-PMo$_{12}$(0)//Pt junction dataset. (a-c) feature spaces for 4, 5 and 6 clusters, respectively. (d-f) Mean Ī-V for 4, 5 and 6 clusters, respectively.

| 4 clusters | ε$_{0-SEL}$ (eV) | ε$_{0-TVS}$ (eV) | 5 clusters | ε$_{0-SEL}$ (eV) | ε$_{0-TVS}$ (eV) | 6 clusters | ε$_{0-SEL}$ (eV) | ε$_{0-TVS}$ (eV) |
|---|---|---|---|---|---|---|---|---|
| c2/4 | 0.69 | 0.65 | c2/5 | 0.69 | 0.61 | c2/6 | 0.69 | 0.65 |
|  |  |  |  |  |  | c3/6 | 0.70 | 0.64 |
| c3/4 | 0.64 | 0.57 | c3/5 | 0.67 | 0.60 | c4/6 | 0.66 | 0.60 |
|  |  |  | c4/5 | 0.60 | 0.55 | c5/6 | 0.60 | 0.54 |
| c4/4 | 0.80 | 0.68 | c5/5 | 0.81 | 0.72 | c6/6 | 0.80 | 0.70 |

**Table S1.** Energy of the MO determined by the SEL model and the TVS method (on the mean Ī-V, Figs. S12, S14 and S15) for the clustering analysis with 4, 5 and 6 clusters. The light gray lines highlight the pertinent clusters.

The same analysis conducted with the dataset of $^{TS}$Au-C6-PMo$_{12}$(I)//Pt junctions (Fig. S13 and Table S2) leads to the same conclusion. The clusters c2/M and c3/M are identical whatever the total number M of clusters (Fig. S13 and Table S2) and the clusters c3/6 and c4/6 (analysis with 6 clusters) are identical and thus this splitting is not useful. Comparing the 4 and 5 clusters analysis, the cluster c4/4 can be decomposed in two clusters c4/5 and c5/5 (Fig. S13) with slightly different parameters (Table S2 and Table 3 in the main text). Thus we also keep an optimized number of 5 clusters for the analysis of the $^{TS}$Au-C6-PMo$_{12}$(I)//Pt junction dataset. This is also consistent for comparison with the analysis of the $^{TS}$Au-C6-PMo$_{12}$(0)//Pt junction dataset.

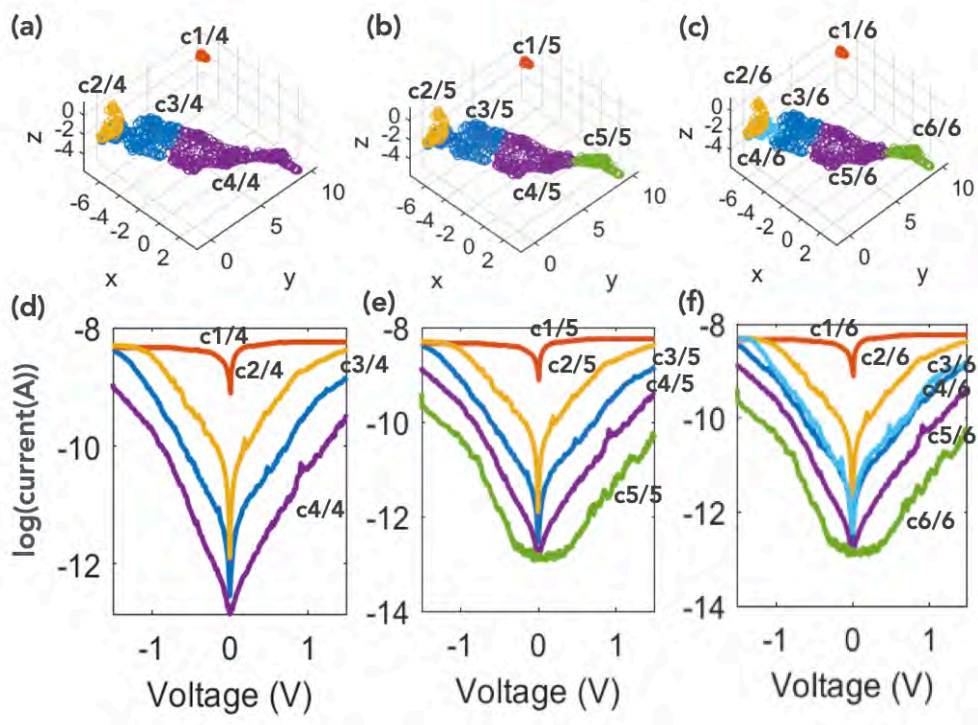

*Figure S13*. $^{TS}$Au-C6-PMo$_{12}$(I)//Pt junction dataset. (a-c) feature spaces for 4, 5 and 6 clusters, respectively. (d-f) Mean Ī-V for 4, 5 and 6 clusters, respectively.

| 4 clusters | $\varepsilon_{0\text{-TVS}}$ (eV) | 5 clusters | $\varepsilon_{0\text{-TVS}}$ (eV) | 6 clusters | $\varepsilon_{0\text{-TVS}}$ (eV) |
|---|---|---|---|---|---|
| c2/4 | 0.41 | c2/5 | 0.43 | c2/6 | 0.41 |
| c3/4 | 0.36 | c3/5 | 0.38 | c3/6 | 0.37 |
|  |  |  |  | c4/6 | 0.38 |
| c4/4 | 0.29 | c4/5 | 0.33 | c5/6 | 0.28 |
|  |  | c5/5 | 0.28 | c6/6 | 0.26 |

*Table S2.* *Energy of the MO determined by the TVS method (on the mean $\bar{I}$-V, Figs. S13, S19 and S20) for the clustering analysis with 4, 5 and 6 clusters. The light gray lines highlight the pertinent clusters.*

***Statistical analysis of the clusters.***

In this section, we present the detailed analysis of the I-V curves in the case of 5 clusters. The cluster 1 (corresponding to I-Vs at the saturation limit of the C-AFM apparatus) is not considered.

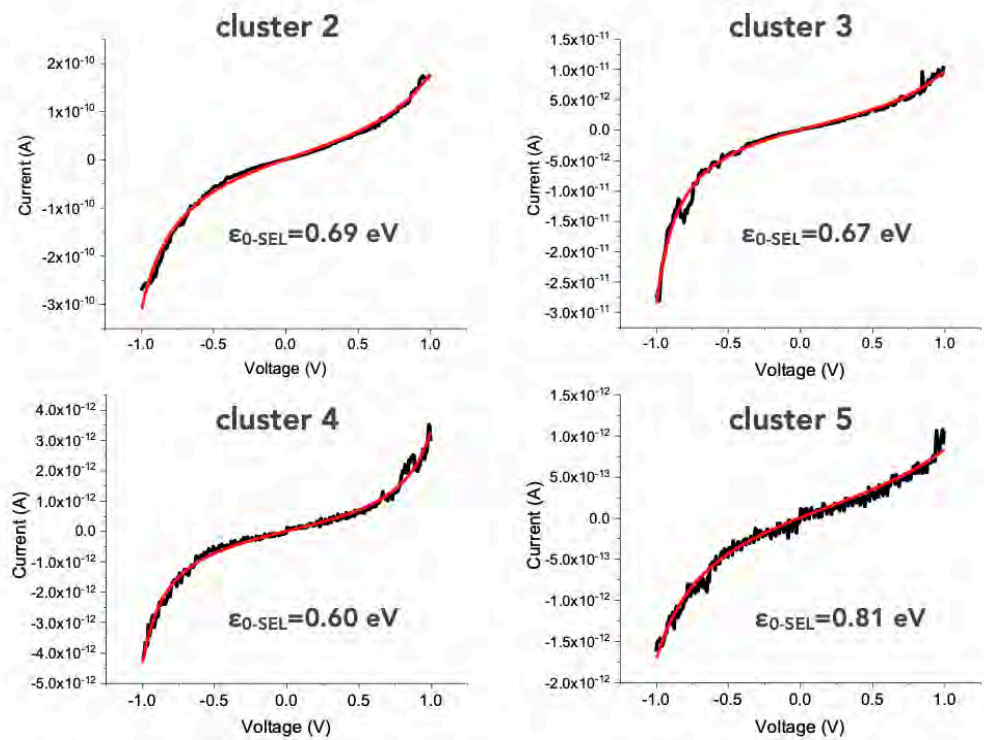

***Figure S14***. *SEL fit (on mean Ī-V) of the four clusters of the $^{TS}$Au-C6/PMo$_{12}$(0)//Pt devices.*

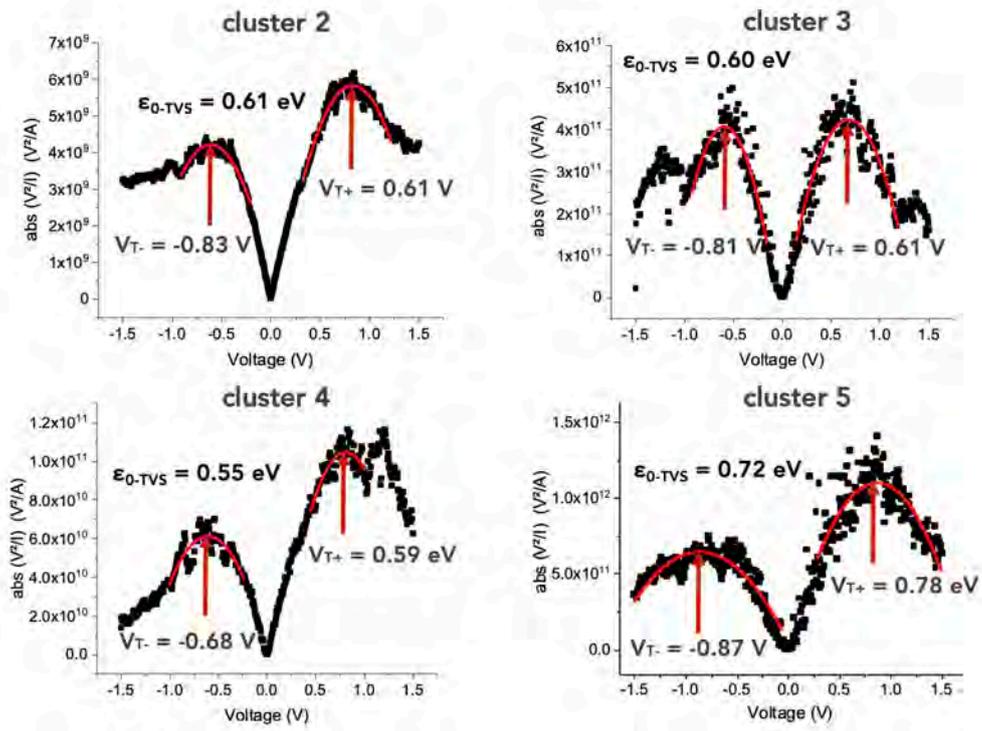

**Figure S15.** TVS analysis of the mean $\bar{I}$-V of the four clusters of the $^{TS}$Au-C6/ PMo$_{12}$(0)//Pt devices (same data as Fig. S14).

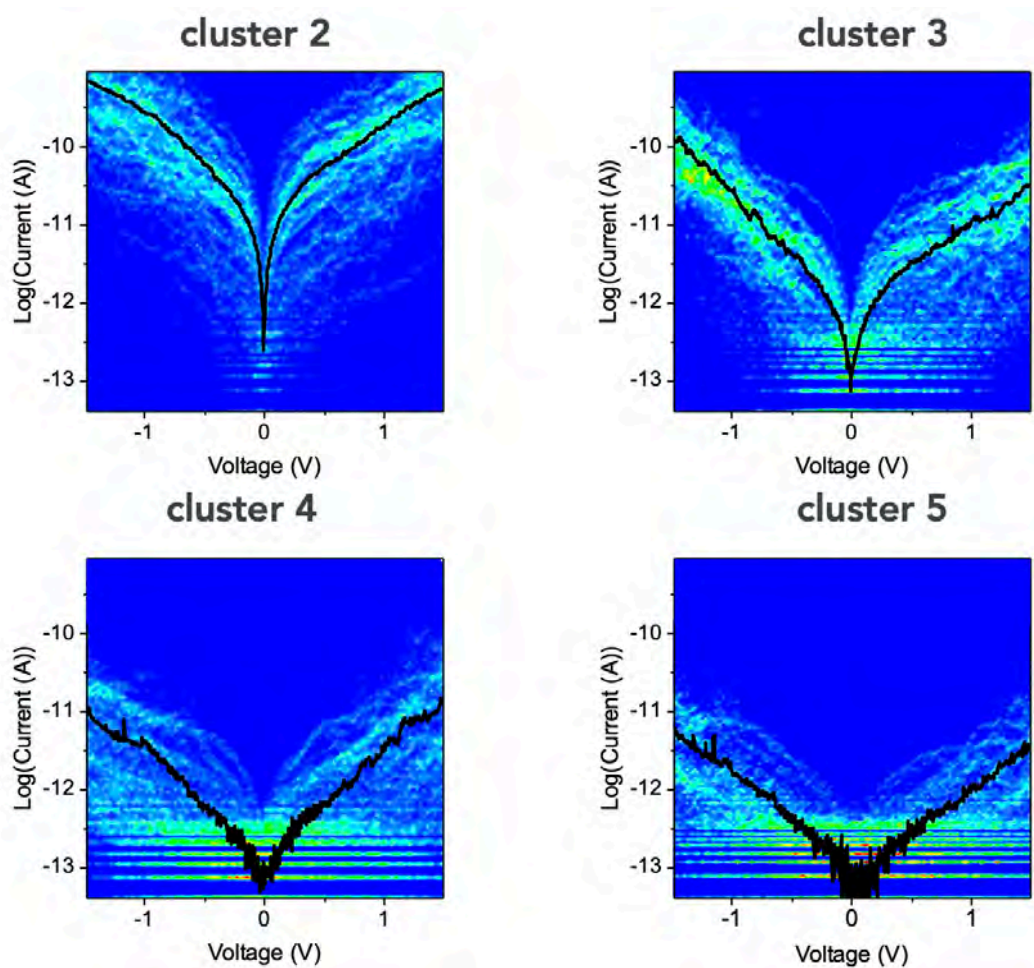

*Figure S16*. Current-voltage (I-V) curves of the $^{TS}$Au-C6/PMo$_{12}$(0)//Pt junctions belonging to each cluster with the mean Ī-V curve (dark lines). The numbers of I-Vs are 134 (cluster 2), 102 (cluster 3), 159 (cluster 4) and 107 (cluster 5).

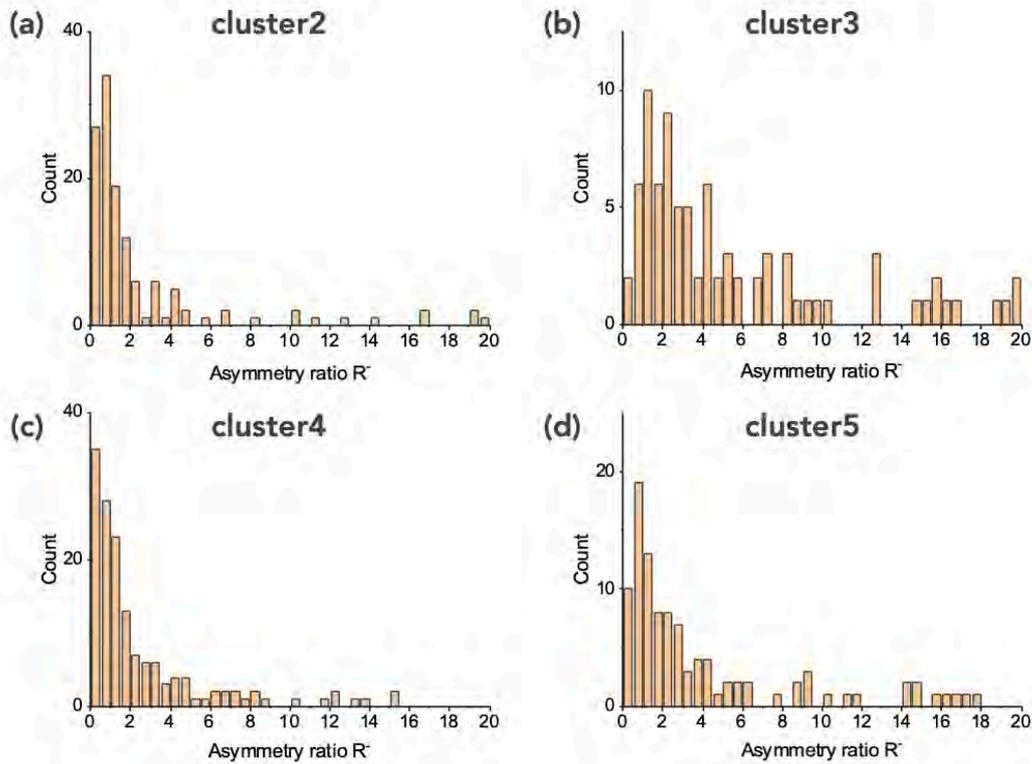

***Figure S17.*** *Histograms of the asymmetry ratios $R^- = I(-1.5V)/I(1.5V)$ calculated from all the individuals I-Vs belonging to each cluster of the $^{TS}$Au-C6/PMo$_{12}$(0)//Pt junctions shown in Fig. S16.*

The histograms of the asymmetry ratio of the $^{TS}$Au-C6/PMo$_{12}$(0)//Pt junctions (Fig. S17) confirm the analysis from the mean Ī-V curves (Table 3 main text) that the cluster 3 contains a majority of I-V curves with an asymmetry ratio larger than 2 (71% of the data), while the 3 other clusters have more than 50% of almost symmetric I-V ($R^-$ < 2 : 72% for the cluster 2, 66 % for the cluster 4 and 50% forthe cluster 5).

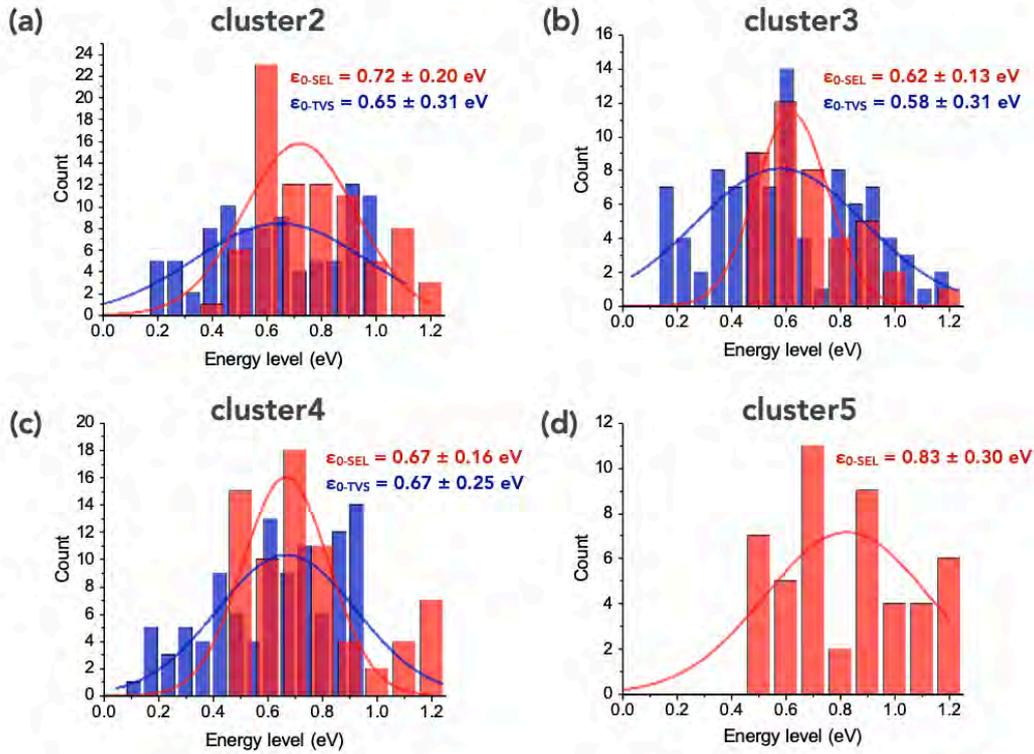

**Figure S18.** *Histograms of the energy levels $\varepsilon_{0\text{-SEL}}$ and $\varepsilon_{0\text{-TVS}}$ obtained with the SEL model and TVS method for all the individuals I-Vs belonging to each cluster of the $^{TS}$Au-C6/PMo$_{12}$(0)//Pt junctions shown in Fig. S16. Solid lines are the fits by a Gaussian distribution with the mean value of the energy level ± standard deviation indicated in the panels.*

The Gaussian fits of distributions of the $\varepsilon_{0\text{-SEL}}$ and $\varepsilon_{0\text{-TVS}}$ values obtained for each cluster (Fig. S18, and Table S3) confirm the value directly obtained from the mean Ī-V curves (Figs. S14, S15) and are compared in Table S3 for convenience. The TVS method for cluster 5 gives a large distribution of value, likely due to the fact that many I-Vs in this cluster are noisy curves at the sensitivity limit of the C-AFM system and they were discarded in this case.

| cluster | TSAu-C6/PMo$_{12}$(0)//Pt | | | | TSAu-C6/PMo$_{12}$(I)//Pt | | | |
|---|---|---|---|---|---|---|---|---|
| | C2 (24.6%) | C3 (18.8%) | C4 (29.2%) | C5 (19.7%) | C2 (12.5%) | C3 (35.7%) | C4 (35.5%) | C5 (13.5%) |
| $\varepsilon_{0\text{-TVS}}$ (eV) mean $\bar{I}$-V | 0.61 | 0.60 | 0.55 | 0.72 | 0.43 | 0.38 | 0.33 | 0.28 |
| $\varepsilon_{0\text{-TVS}}$ (eV) histogram | 0.65 | 0.58 | 0.67 | n.a. | 0.45 | 0.34 | 0.35 | n.a. |
| $\varepsilon_{0\text{-SEL}}$ (eV) mean $\bar{I}$-V | 0.69 | 0.67 | 0.60 | 0.81 | n.a. | | | |
| $\varepsilon_{0\text{-SEL}}$ (eV) histogram | 0.72 | 0.62 | 0.67 | 0.83 | | | | |

*Table S3*. Comparison of the energy level $\varepsilon_0$ for the TVS method and the SEL model deduced from the mean $\bar{I}$-V curves (Figs. S14, S15, S19) and from Gaussian fits of the histograms belonging to the different clusters (Figs. S18 and S23).

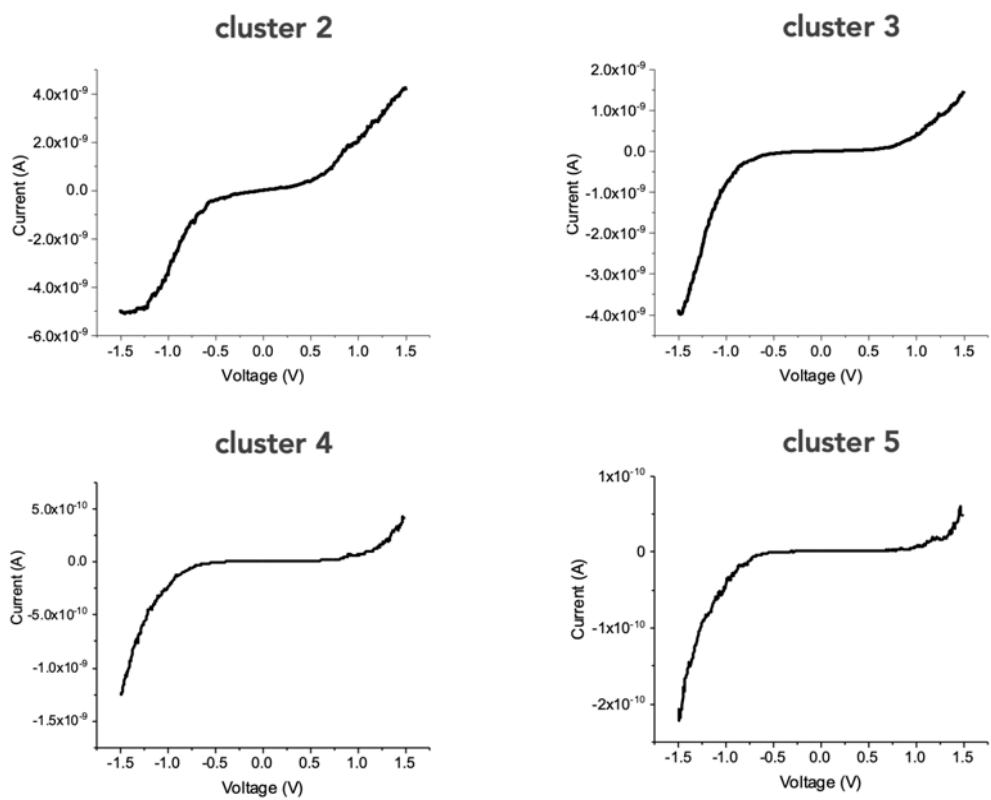

*Figure S19*. Mean $\bar{I}$-V of the four clusters of the TSAu-C6/PMo$_{12}$(I)//Pt devices.

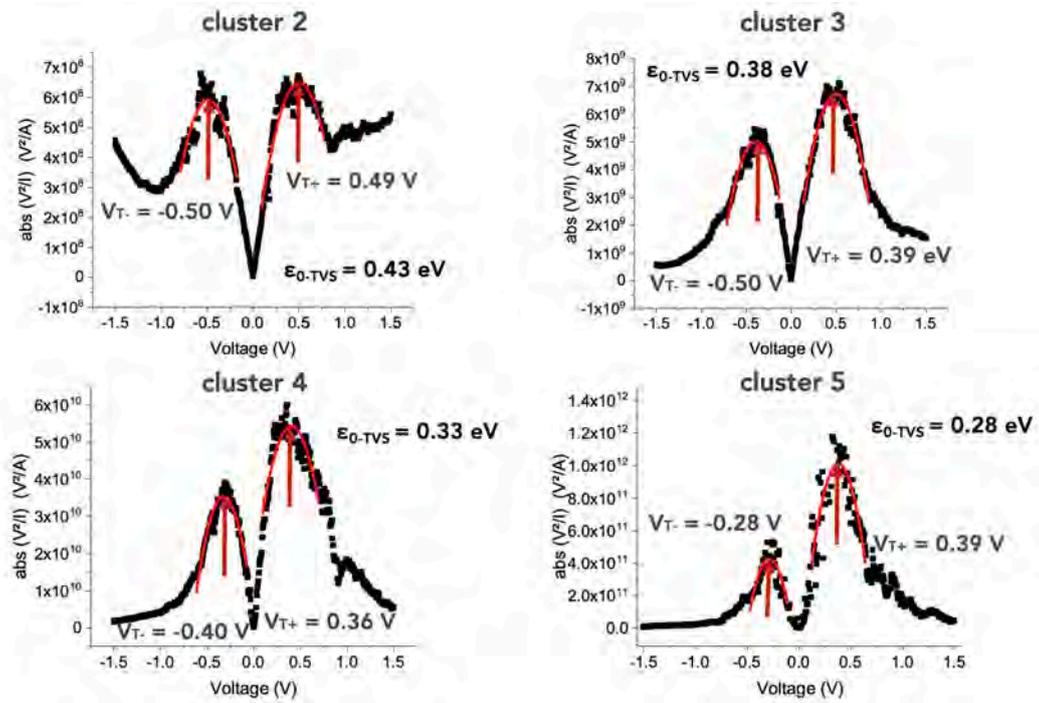

**Figure S20**. TVS analysis of the mean Ī-V of the four clusters of the $^{TS}$Au-C6/ PMo$_{12}$(I)//Pt devices (same data as Fig. S19).

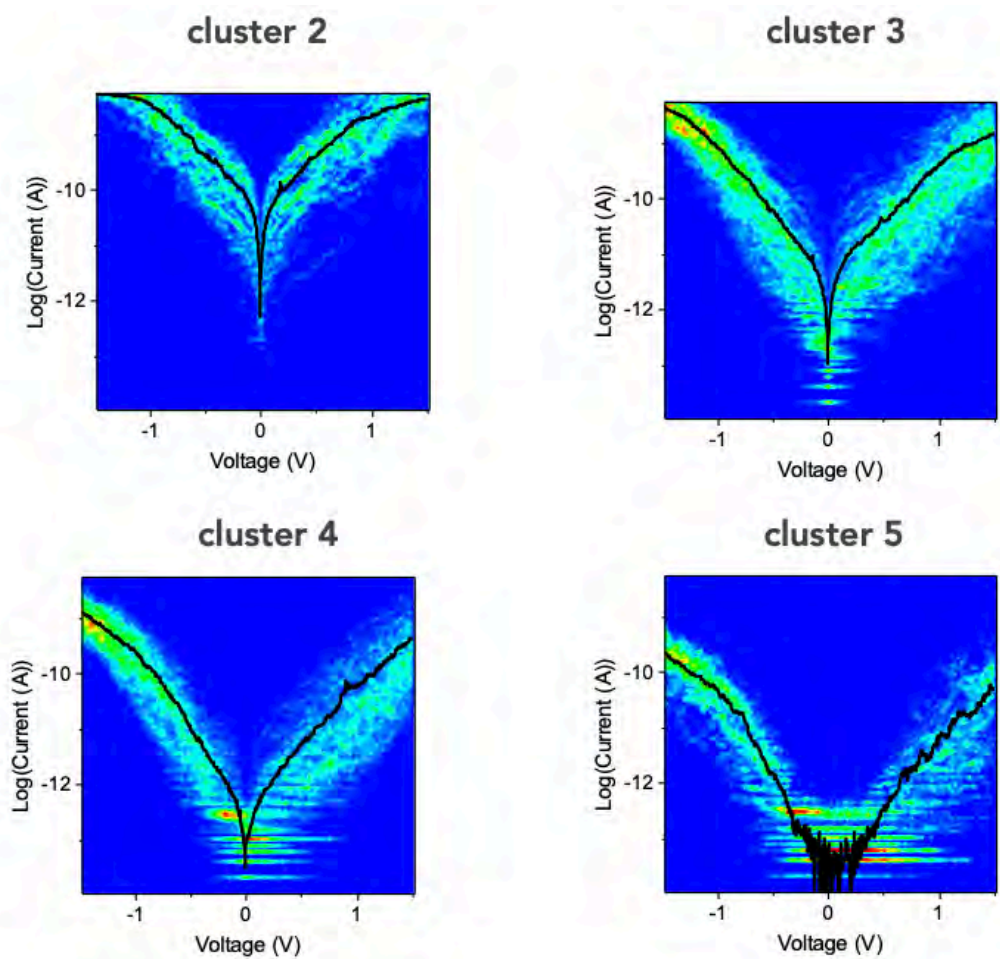

**Figure S21**. Current-voltage (I-V) curves of the $^{TS}$Au-C6/PMo$_{12}$(I)//Pt junctions belonging to each cluster with the mean Ī-V curve (in red). The numbers of I-Vs are 75 (cluster 2), 214 (cluster 3), 213 (cluster 4) and 81 (cluster 5).

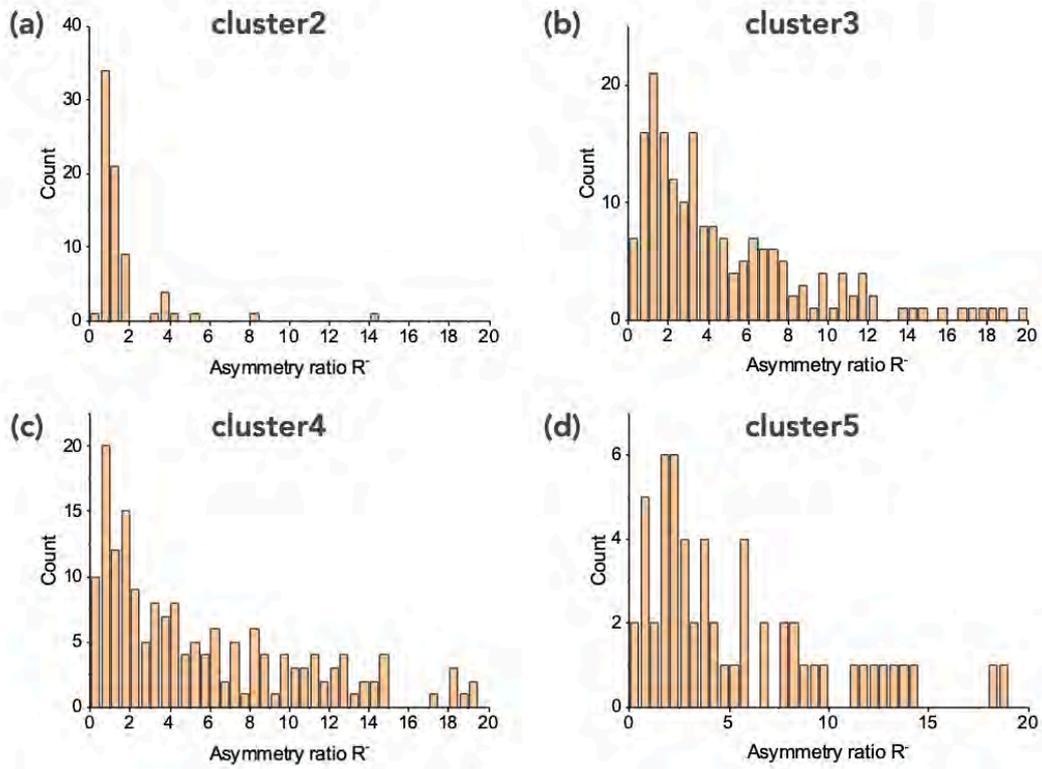

***Figure S22.*** *Histograms of the asymmetry ratios $R^- = I(-1.5V)/I(1.5V)$ calculated from all the individuals I-Vs belonging to each cluster of the $^{TS}Au\text{-}C6/PMo_{12}(I)//Pt$ junctions shown in Fig. S21.*

The histograms of the asymmetry ratio (Fig. S22) confirm the analysis from the mean Ī-V curves (Table 3 main text) that all the clusters for $^{TS}Au\text{-}C6/PMo_{12}(I)//Pt$ junctions contain a majority of I-V (≈ 70% of the data) with an asymmetry ratio larger than 2 (68% for the cluster 3, 67 % for the cluster 4 and 74% for the cluster 5). Note that, as for the analysis from the mean Ī-V curve, the ratio for the cluster 2 is not meaningful since almost all the I-Vs saturate (compliance of the preamplifier of the C-AFM apparatus) at voltage ± 1.5 V( Fig. S21).

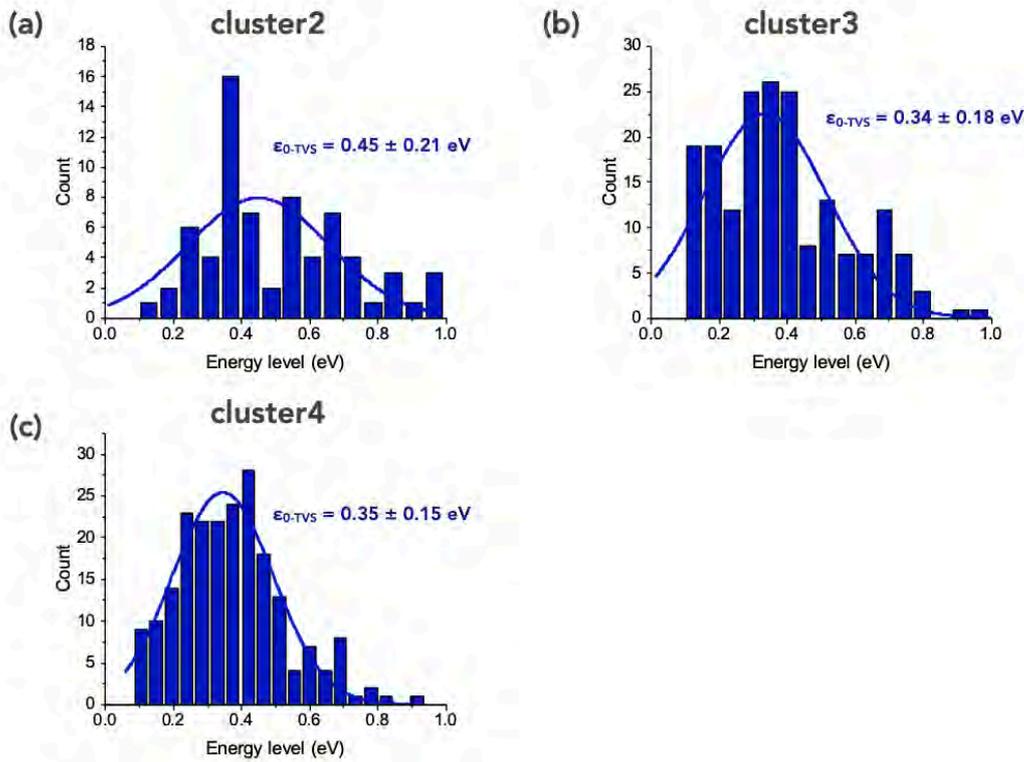

*Figure S23.* Histograms of the energy levels $\varepsilon_{0\text{-}TVS}$ obtained with TVS method for all the individuals I-Vs belonging to each cluster of the $^{TS}$Au-C6/PMo$_{12}$(I)//Pt junctions shown in Fig. S21.

The histograms of $\varepsilon_{0\text{-}TVS}$ values (TVS analysis) conducted for each cluster (Fig. S23, and Table S3) confirm the value directly obtained from the mean $\bar{I}$-V curve (Figs. S19, S20) and are compared in Table S3 for convenience. The TVS method for the cluster 5 gives a large distribution of value, likely due to the fact that many I-Vs in this cluster are noisy curves at the sensitivity limit of the C-AFM system and was discarded in this case.

**Section 8. References samples, C6 SAMs.**

Figure S24a shows the I-V dataset (415 I-Vs) measured by C-AFM on $^{TS}$Au-C6//Pt samples. The fit by the SEL model and the TVS method on the mean $\bar{I}$-V give $\varepsilon_{0\text{-}SEL}$

= 0.85 eV (Fig. S24b) and $\varepsilon_{0\text{-TVS}}$ = 0.72 eV (Fig. S24c). The statistical analysis of the complete data set gives almost the same mean value of ~ 0.9 eV for both methods (Fig. S24d). This latter value is in good agreement with the energy position of the LUMO for alkyl chains on Au.[20, 21]

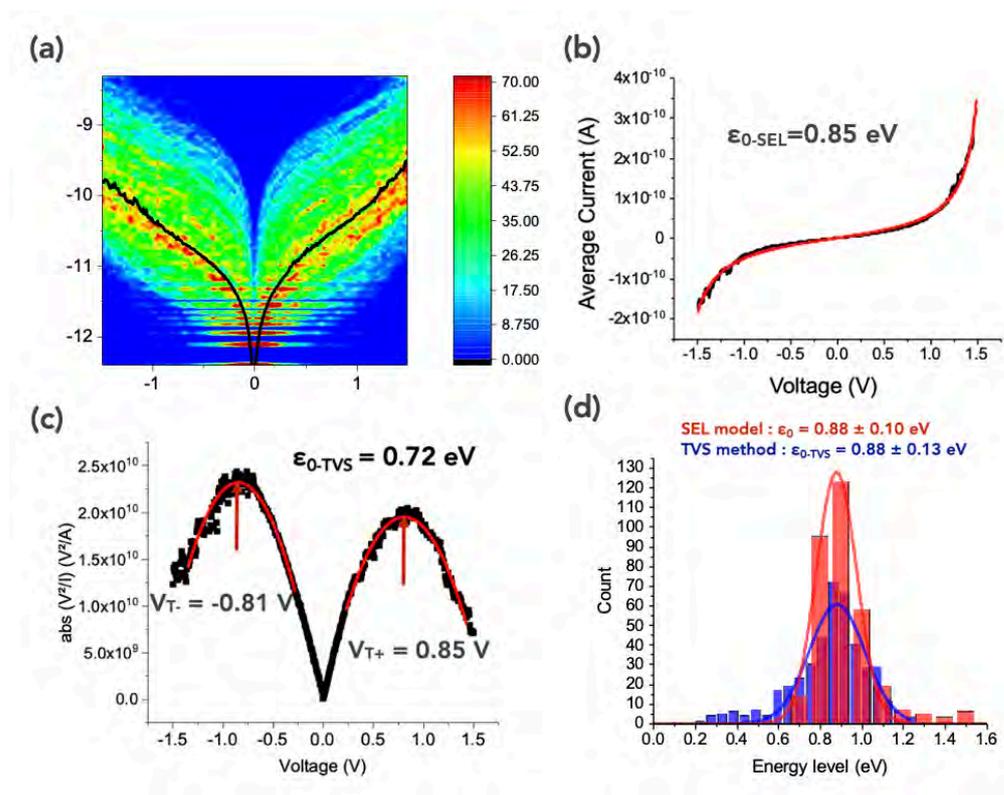

*Figure S24.* (a) 2D histogram of 415 I-Vs of $^{TS}$Au-C6//Pt junction, and mean current $\bar{I}$-V curve (dark line), (b) typical fit of the SEL model and (c) TVS method on the mean $\bar{I}$-V, (d) statistical distribution of the energy level and fit with a Gaussian distribution.